\documentclass{article}
\usepackage{graphicx} 
\usepackage{amsmath,amssymb,amsfonts}
\usepackage{soul}
\usepackage[table]{xcolor}
\usepackage{graphicx}
\usepackage{subcaption}
\usepackage{multirow}
\usepackage{mathrsfs}
\usepackage{hyperref}
\usepackage{tabularx}
\usepackage{amsthm}
\usepackage{colortbl}
\usepackage{pdflscape}
\usepackage[linesnumbered,ruled,vlined]{algorithm2e}
\DeclareMathOperator*{\argmax}{argmax}
\newcommand{\model}{\textsc{Cuttana}\ }
\newcommand{\fennel}{\textsc{Fennel}\ }
\newcommand{\heistream}{\textsc{HeiStream}\ }
\newcommand{\ldg}{\textsc{Ldg}\ }

\usepackage{lastpage}

\usepackage{mathtools}

\usepackage[normalem]{ulem}
\useunder{\uline}{\ul}{}

\usepackage{textcomp}
\usepackage{xcolor}
\usepackage{caption}
\title{CUTTANA: Scalable Graph Partitioning for Faster Distributed Graph Databases and Analytics}
\author{Milad Rezaei Hajidehi\\ \texttt{miladrzh@cs.ubc.ca}
\and
Sraavan Sridhar\\ \texttt{sraavan@student.ubc.ca}
\and
Margo Seltzer\\ \texttt{mseltzer@cs.ubc.ca}
}
\date{} 

\begin{document}

\maketitle

\section{Introduction}


\textbf{Ubiquity and massive growth of real-world networks sparked the applications of distributed graph processing.}
A graph is a common data model that can represent complex relationships between real-world entities in myriad domains such as social networks, the World Wide Web, finance, fraud detection, transportation, and biological networks \cite{sahu2017ubiquity}. 
In practice, many of these graphs are sufficiently large to exceed the memory of a single machine, posing performance challenges for single-node solutions. Using distributed systems with increased memory and parallelism enables high performance for large graph processing. The ubiquity and growth of real-world graphs motivated the development of distributed graph processing solutions for various applications such as graph analytics \cite{gonzalez2012powergraph,chen2019powerlyra,GraphX,fan2021graphscope,li2019topox,yan2020g,iyer2021tegra,wang2022scaleg,chen2023khuzdul}, graph databases \cite{li2022bytegraph,buragohain2020a1,TitanDB,JanusGraph}, and graph neural networks (GNN) \cite{md2021distgnn,wang2021gnnadvisor,vatter2023evolution,zheng2020distdgl,gandhi2021p3}. 



\begin{figure}[t]
\centering
\includegraphics[width=0.8\linewidth]{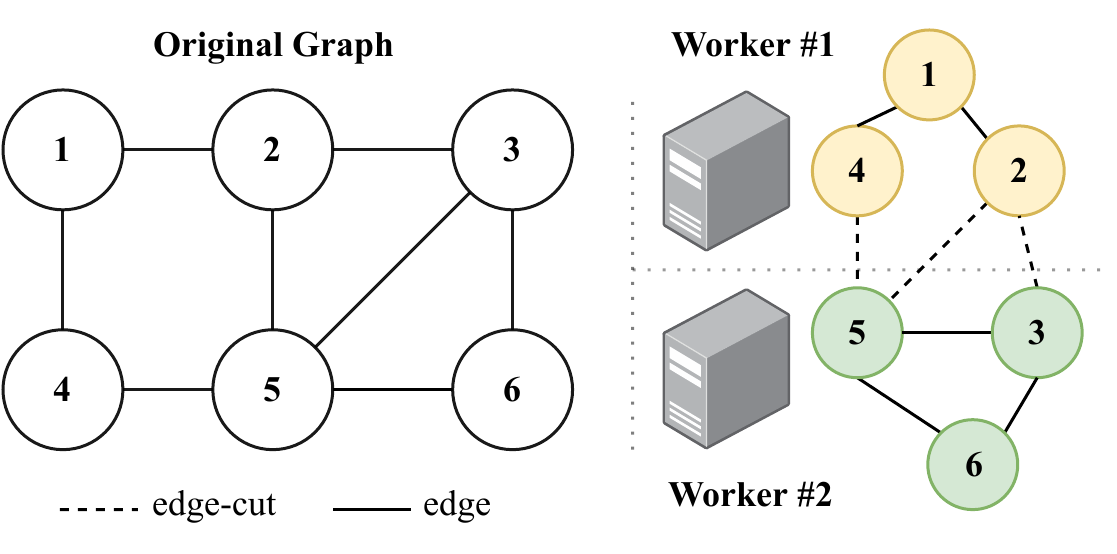}
\centering
\captionsetup{justification=centering,font=small} 
\caption{Partitioning a graph for two workers. Transferring data through edge-cuts (dotted edges) requires network calls.}
\label{fig:graphpartition}
\end{figure}

\begin{figure}[t]
\centering
\includegraphics[width=0.93\linewidth]{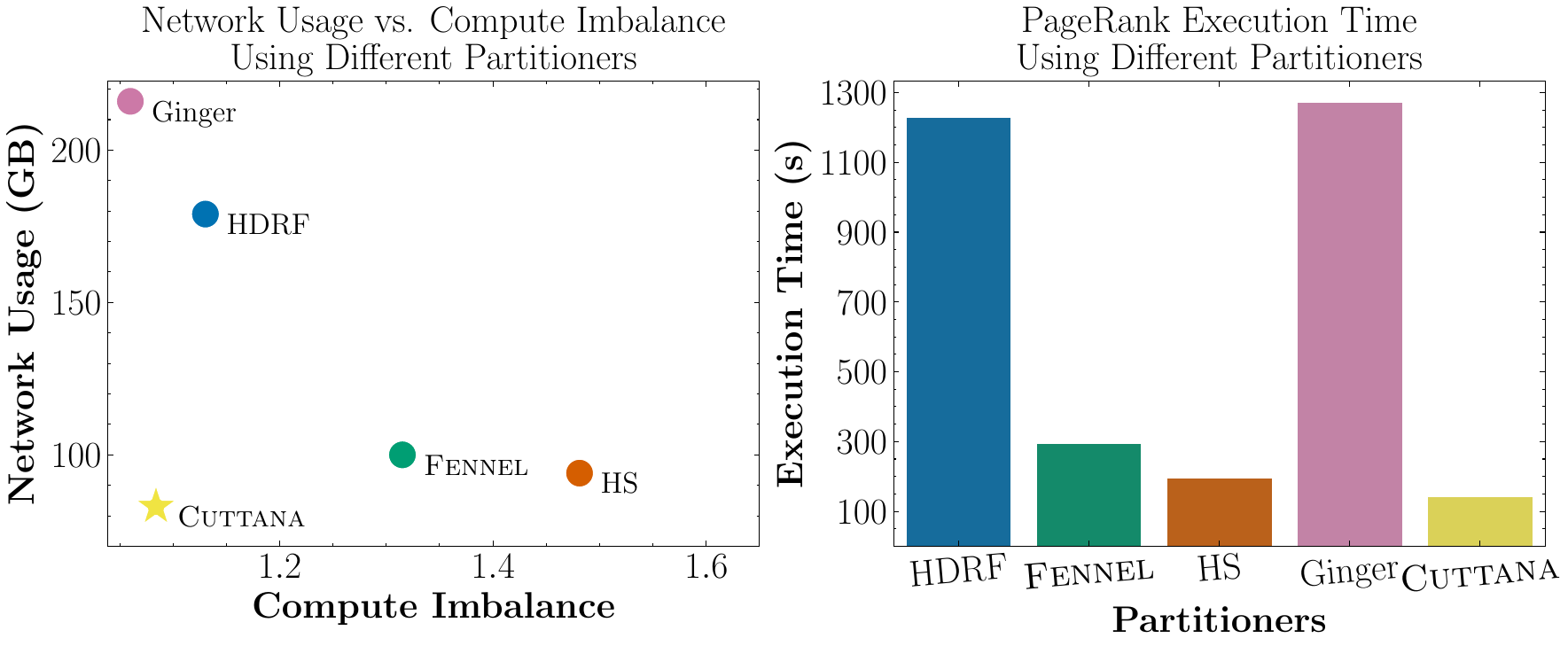}
\centering
\captionsetup{font=small} 
\caption{An example of partitioning's effect on network usage, compute imbalance, and total time of PageRank on the UK07 dataset. \model improved PageRank execution time by more than 150s (52\%) relative to \textsc{Fennel}\cite{tsourakakis2014fennel} and (52s) 27\% relative to \textsc{HeiStream} \cite{faraj2022buffered}, while reducing partitioning time more than 50\%.}
\label{fig:crown}
\end{figure}

\textbf{Graph partitioning affects the performance of distributed graph processing.} The first step of any distributed graph processing application is to partition the graph into disjoint subgraphs and distribute them to worker machines. Unlike traditional distributed applications such as map-reduce, graph processing workloads exhibit many interactions among partitions \cite{mccune2015thinking}. For example, in PageRank, the rank of a vertex is calculated based on the rank of its neighbors in each iteration. To achieve high-quality partitioning, the number of edges that have vertices assigned to different machines (i.e., edge-cuts) should be minimized, since exchanging data along those edges incurs network overhead (Figure \ref{fig:graphpartition}).  Another aspect of good partitioning is assigning equal-sized partitions to workers to avoid stragglers. Figure \ref{fig:crown} shows that network overhead can be more than 100GB for a 16-worker PageRank computation on the UK07 dataset and the graph partitioning algorithm has a significant effect on network usage, worker imbalance, and execution time. 

\textbf{Partitioning large graphs is hard and memory-bound.}
The problem of balanced graph partitioning is $\mathcal{NP}$-hard \cite{garey1974some}. However, many domains other than distributed graph processing (e.g., VLSI design \cite{ccatalyurek2023more} and causal inference \cite{awadelkarim2020prioritized}) demand high-quality partitioning. As a result, many heuristic solutions  exist~\cite{karypis1998fast,sanders2011engineering,zhang2017graph,mayer2021hybrid,martella2017spinner,ccatalyurek2023more}.
However, most of these solutions fail when partitioning graphs larger than the main memory. For example, \textsc{Metis} \cite{karypis1998fast}, long the gold standard for graph partitioning, is unable to partition the Twitter or Web graphs \cite{mayer2021hybrid,margo2017sorting}, leading to the development of various streaming partitioners for massive graphs \cite{faraj2022buffered,chen2019powerlyra,stanton2012streaming,tsourakakis2014fennel,mayer2018adwise,pacaci2019experimental}. 

\textbf{Streaming partitioners are scalable but low-quality.}
Streaming solutions make partitioning decisions by reading vertices or edges one by one and assigning them to partitions based on a scoring function. The score is calculated from minimal summarized information about the vertices/edges already assigned, the current vertex/edge, and the partition sizes. There are two types of partitioners: vertex-partitioners  (edge-cut partitioners) \cite{stanton2012streaming,tsourakakis2014fennel,faraj2022buffered}, which read a stream of vertices and their neighbors and assign each vertex to a partition, and edge partitioners (vertex-cut partitioners) \cite{petroni2015hdrf,chen2019powerlyra,fan2023application}, which read a stream of edges and assign each to a partition. In an experimental study, Pacaci and Ozsu reported that edge-cut partitioners yield lower network overhead but greater worker imbalance \cite{pacaci2019experimental}. Analyses demonstrate the inferior quality of streaming partitioners for small-to-mid scale graphs relative to in-memory partitioners, which is unsurprising given their limited view of the original graph \cite{tsourakakis2014fennel,awadelkarim2020prioritized}.

\textbf{\model is a high-quality, scalable partitioner}, designed to have the scalability of streaming solutions while providing better partitioning quality. We studied existing streaming edge-cut partitioners and found three major limitations.
1) They prematurely assign vertices when the data needed to calculate an accurate scoring function is not available. 
2) They never change vertex assignments, even though, over time, the algorithm gains information about the graph and its structure. 
3) The significant worker imbalance when using edge-cut partitioners overshadows their network overhead superiority.

We solve the first problem by introducing \textit{score-based dynamic buffering}. We buffer vertices based on the knowledge we have about their neighborhood and avoid premature partitioning when insufficient data is available. However, if done naively, buffering can result in storing the entire graph in memory, which is obviously not scalable. We solve the second problem by providing a mechanism to move and exchange vertices between partitions to enhance the partitioning quality at the end of the streaming phase. Determining which moves enhance quality requires saving the neighborhood for each vertex and is also impossible (due to memory constraints). Also, the moving phase can be time-consuming due to the large number of possible moves. We introduce a \textit{coarsening} strategy and a theoretically efficient \textit{refinement} algorithm to find the best moves, enabling fast and coarse-grained improvement of partitioning quality. We show that the huge edge imbalance in existing edge-cut partitioners is the cause of worker imbalance in analytics.
We solve the third problem by modeling and satisfying an edge-balance condition using an edge-based score function and our refinement algorithm. Finally, to minimize the potential time overhead caused by buffering and refinement, we provide a parallel implementation that yields nearly the same partitioning time for massive graphs compared to streaming solutions, while offering better partitioning quality. 

\textbf{Our contributions} are as follows.

\begin{itemize}
    \item{We present a scalable, buffered streaming partitioning model to effectively use main memory to avoid premature vertex assignment. This model can be applied to any existing streaming partitioner to increase its quality.}
    \item We introduce a novel \textit{coarsening} and \textit{refinement} technique that receives the output of a streaming partitioner and improves it to reach a ``maximal'' quality state. This algorithm is theoretically efficient and independent of the size of the graph. 
    \item We leverage unused cores via a parallel implementation, providing rapid partitioning speed. 
    \item Through experimental analysis, we show \textsc{Cuttana}'s superiority relative to existing edge-cut partitioners. We also demonstrate the edge imbalance of existing partitioners, which is often overlooked in the literature. 
    \item We show the effect of \model partitioning quality improvement in the execution time of distributed graph analytics. Overall \model can improve the runtime performance of graph analytics by up to 59\% and is the best partitioner in most scenarios. 
    \item We show the effect of \model partitioning quality improvement in the query throughput of distributed graph databases. \model can improve the throughput of the JanusGraph distributed graph database by up to 23\% over the best existing graph partitioner in the standard LDBC social network benchmark. 
\end{itemize}

\newcommand{\V}[0]{\mathcal{V}}
\newcommand{\E}[0]{\mathcal{E}}
\newcommand{\K}[0]{\mathcal{K}}
\newcommand{\p}[0]{\mathcal{P}}
\newcommand{\N}[0]{\mathcal{N}}
\newcommand{\D}[0]{\mathcal{D}}
\newcommand{\slack}[0]{\epsilon}
\newcommand{\eclamb}[0]{\lambda_{EC}}
\newcommand{\cvlamb}[0]{\lambda_{CV}}

\newtheorem{example}{Example}

\section{Background}\label{sec:bg}

\textbf{Formal definition of vertex partitioning problem.} Given a graph $G=\langle V,E \rangle$, the  $\K$-way vertex-balanced graph partitioning problem is to assign vertices to the disjoint sets $\V_1,\V_2,\ldots,\V_\K$ such that $\bigcup_{i=1}^{\K} \V_i=V$ and the $\V_i$ satisfy the balance condition:

 \begin{equation}
     |\V_i| \leq (1 + \slack). \frac{|V|}{\K} \ \  \ \ (1 \leq i \leq \K)
     \label{eq:balance}
 \end{equation}

The $\slack \geq 0 $ is the balance slack parameter that constrains how imbalanced the partitions can be. The balance condition can also be defined based on the number of edges in a partition. With $\N(v)$ representing the set of neighbors for vertex $v$, we define the edge-balance condition for vertex partitioning as:

\begin{equation}
\sum_{v \in \V_i}|\N(v)| \leq (1 + \slack) \cdot \frac{2 \cdot |E|}{\K} \quad (1 \leq i \leq \K)
\label{eq:edgebalance}
\end{equation}

\textbf{Optimization objectives.}
The quality metrics for graph partitioning are based on minimizing the interdependency of partitions. A common metric is \textit{edge-cut}, the number of edges whose endpoints are in different partitions. Given $\p \colon V \to \mathbb{N}_{\leq\K}$, the function that returns the partition ID to which a vertex is assigned, the normalized number of edge-cuts is: 

\begin{equation}
 \eclamb = \frac{|\{ \langle x,y \rangle \in E | \p(x) \neq \p(y) \}|}{|E|}
     \label{eq:ec}
 \end{equation}
 
Minimizing edge-cuts is equivalent to minimizing network cost, since whenever a vertex requires data from a neighboring vertex in a different partition, the two corresponding workers must transmit the data over the network. 

A common optimization in bulk synchronous systems, mostly in analytic workloads, is \textit{sender-side aggregation} \cite{bourse2014balanced,pacaci2019experimental,mccune2015thinking}. In these systems, the workload is iterative (e.g., PageRank iteration) and at the end of each iteration, if multiple vertices in the same worker are connected to the same vertex in a different worker, the neighboring vertex sends the data once. This causes all of the edges between the neighboring vertex and the vertices in the first worker to need only a single network message. In Figure \ref{fig:graphpartition}, vertices 2 and 5 can benefit from this optimization. \textit{Communication-volume} is the metric that models the network cost of such systems. Given $\D \colon V \to \mathbb{N}_{\leq\K}$, a function that returns the number of partitions in which a given vertex has neighbors, excluding its own partition ($\p(v)$), the normalized communication volume is:

\begin{equation}
 \cvlamb = \frac{\sum_{u\in V} \D(u)}{\K|V|}
     \label{eq:cv}
 \end{equation}

Generally, edge-cut is a metric that models network traffic for asynchronous systems such as graph databases, and communication volume is a metric that models network traffic for synchronous systems \cite{pacaci2019experimental}. 

\textbf{General streaming model for edge-cut graph partitioning.} At each iteration $t$ ($1\leq t\leq |V|$) where we read the $t$\textsuperscript{th} vertex in the stream, a streaming edge-cut partitioner reads the vertex $u_t$ and its neighbours $\N(u_t)$ and assigns $u_t$ to one of the partitions. The assignment is based on evaluating a \textit{score} function for each partition based on $u_t$, $\N(u_t)$, and the state of each partition in the $t$\textsuperscript{th} iteration ($\V_i^t$). A general model for assignment of $u_t$ is:

\begin{equation}
\argmax_{1\leq i\leq \K} \left[\mathbf{h}(|\V_i^t\cap\N(u_t)|) - \mathbf{g}(|\V_i^t|)\right]
\label{eq:score1}    
\end{equation}
where $\mathbf{h}$ biases assigning the vertex to the partition that contains the greatest number of neighbors, thus minimizing the number of edge cuts, and $\mathbf{g}$ is the penalty term for the current size of the partition to satisfy balance constraints, thus encouraging equal partition growth. This heuristic is at the core of many edge-cut partitioners, and variants of Equation \ref{eq:score1} can be found in them \cite{tsourakakis2014fennel,stanton2012streaming,awadelkarim2020prioritized,nishimura2013restreaming,huang2016leopard,fan2020incrementalization}.


\newtheorem{theorem}{Theorem}
\newtheorem{proposition}{Proposition}
\newtheorem{corollary}{Corollary}
\newtheorem{lemma}{Lemma}
\theoremstyle{definition}
\newtheorem{definition}{Definition}
\theoremstyle{definition}

\section{\model Algorithm}

 \textbf{Scope.} \textsc{Cuttana} is a vertex partitioner that operates on a static snapshot
of a graph and is designed to improve workload latency and combined workload/partitioning latency for jobs on distributed vertex-centric systems. We designed \textsc{Cuttana} so that it can be executed on commodity machines commonly used for distributed processing in the cloud (concerning their memory constraints). The main focus of \textsc{Cuttana} is on massive graphs (e.g., billion-scale graphs) for which in-memory partitioners (e.g., \textsc{Metis}) fail.

\textbf{Overview.} \model is a two-phase partitioner. The first phase is a streaming partitioner with delayed placement that creates an initial partitioning of the graph. The second phase is the refinement of the initial partitioning. We move vertices among the partitions to increase the partitioning quality (e.g., reducing edge-cuts or communication volume) while maintaining the balance condition.

The delayed placement in the first phase is incorporated into a streaming algorithm by means of a buffered streaming model. This model enables any classic streaming partitioner to delay the assignment of a vertex whenever necessary; we discuss this in Section \ref{sec:p1}. In Section \ref{sec:p2}, we explain the challenges of refinement and how our solution addresses them. 
Finally, in Section \ref{sec:parallel}, we explain how we reduce the time overhead introduced by buffering and refinement.

\subsection{Phase 1: Prioritized Buffered Streaming}\label{sec:p1}


\textbf{Premature assignments: a problem in streaming partitioners.}
The primary intuition behind streaming partitioners is to assign each vertex to the partition containing the greatest number of neighboring vertices. The corresponding term for this greedy assignment in Equation \ref{eq:score1} is $|\V_i^t\cap\N(u_t)|$. However, a partitioner frequently encounters a vertex for which many, or even all, of its neighbors are not yet assigned. We call such assignments \emph{premature}. Mathematically, premature assignments happen when partitioning $u_t$ and the number of assigned neighbors, $\sum_{1 \leq i \leq \K} |\N(u_t) \cap \V_i |$, is small or zero. 
Without adequate information about the assignment of neighboring vertices, the assignment of $u_t$ causes random/low-quality assignment and increases the number of edge-cuts. 


\textbf{Challenges of avoiding premature assignments.}
A simple fix for premature assignment is delaying the assignment of these vertices and prioritizing the assignment of vertices with more already-assigned neighbors. As we partition more vertices, more information becomes available for vertices that were subjected to premature assignment. However, delayed partitioning requires storing all delayed vertices and their neighbors in a buffer, because the streaming phase reads the input file only once, and, after reading a vertex and its neighbors, they are no longer accessible unless explicitly stored. Storing vertices and their neighbors requires $O(E)$ space for the buffer. Limiting the buffer size to a constant, according to the system's available memory, can be a solution, but it requires that the number of buffered vertices be significantly smaller than the size of the entire graph, since high-degree vertices can occupy a significant portion of the buffer. Hence, the key design challenge is determining which vertices to buffer, when to evict them, and how to manage buffering and prioritizing efficiently.




\begin{figure}[t]
\centering
\includegraphics[width=0.9\linewidth]{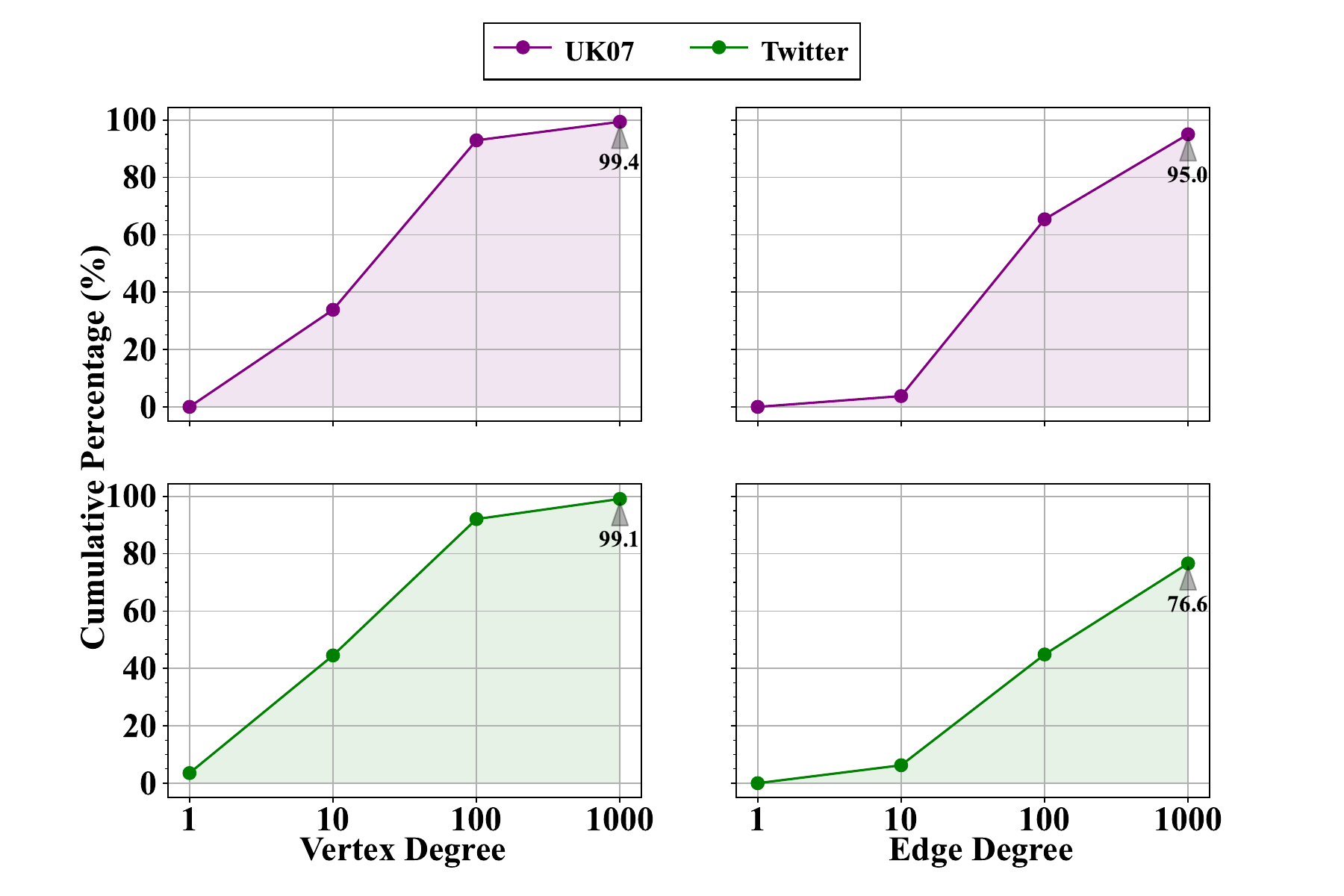}
\centering
\captionsetup{font=small} 
\caption{In large power-law graphs, the majority of vertices have low degrees ($\leq$ 1000) (left charts), and the majority of edges have at least one low-degree endpoint (right charts), even though the maximum degree in these networks exceeds a million.}
\label{fig:powerlaw}
\end{figure}

\textbf{Key Finding: buffering low-degree vertices is sufficient.}
Given that our buffer must hold all the neighbors of the vertices for which we delay assignment, the degrees of the delayed vertices determine the capacity of our buffer. Thus, we should prefer to buffer low-degree vertices (practically, those with fewer than 100 neighbors) over high-degree vertices. Fortunately, this decision proves to be advantageous from a quality perspective as well. First, most large networks exhibit a power-law degree distribution \cite{petroni2015hdrf,albert2000error,kleinberg1999web}, in which the majority of vertices have a low degree, and the majority of edges have at least one low-degree side \cite{mayer2021hybrid}. 
Second, premature assignment is more likely for a low-degree vertex than a high-degree one, because the probability of having zero or a low number of assigned neighbors is inversely proportional to the number of neighbors (Theorem \ref{theorem:one}). Figure \ref{fig:powerlaw} illustrates the first point by showing the cumulative percentage of vertices and edges per degree on two real-world, massive graphs from different domains (web and social). We define the \textit{edge degree} as the minimum degree of its endpoints.

 \begin{theorem}
 \label{theorem:one}
In a streaming partitioner, the degree of a vertex is inversely proportional to the probability of the vertex being partitioned without knowledge of its neighbors.
\end{theorem}
\begin{proof}
When placing the $t$\textsuperscript{th} vertex, $v_t$, with degree $d = |\N(v_t)|$, it will be partitioned with no knowledge if all of its $d$ neighbors come after it in the stream. There are $|V| - t$ such positions, so there are $|V| - t \choose d$ possible orderings. All of the possible ways to place these $d$ vertices in the stream is  $|V| - 1 \choose d$. Thus, the probability that $v_t$ is partitioned with no knowledge is the ratio of the number of orderings in which all the neighbors come after $v_t$ to the total number of possible orderings of $v_t$'s neighbors:

 $$P = \frac{{|V| - t \choose d}}{{|V| - 1 \choose d}} = \frac{(|V| - 1 - d)!(|V| - t)!}{(|V| - 1)!(|V|-t-d)!} = \prod_{i=1}^{t-1} \frac{|V|-d-i}{|V|-i}$$

As $d$ decreases, the numerator increases, yielding higher probability for low-degree vertices. Since $|V| >> d$ for low-degree vertices, the fraction is close to 1, and $P$ can be high, even for large $t$.   
\end{proof}


Therefore, when buffering in a streaming partitioner, the assignment of high-degree vertices remains unchanged, and buffering allows for better assignment of low-degree vertices, which account for the majority of edges in the graph. Finally, buffering low-degree vertices drastically reduces the overhead of buffering and leaves room to buffer more vertices. Hence, buffering only low-degree vertices is a practical performance decision that also enhances partitioning quality.

\begin{algorithm}[!t]
\caption{\textsc{Cuttana}'s First Phase with User-Defined Buffer Score and Partitioning Score Functions}
\SetKwFunction{findBestPartition}{findBestPartition}
\SetKwFunction{updateBuffer}{updateBuffer}
\SetKwFunction{bufferScore}{bufferScore}
\SetKwFunction{partitionVertex}{partitionVertex}
\SetKwFunction{readLine}{readLine}

\SetAlgoNlRelativeSize{0}
\SetAlgoNlRelativeSize{-1}
\SetNlSty{textbf}{(}{)}

\SetAlgoNlRelativeSize{-2}

\SetNlSty{}{}{}

\KwData{Graph File: $F$, Degree Threshold: $D_{\text{max}}$, \newline Vertex Count: $|V|$, Queue Size: $max\_qsize$}

\SetKwProg{function}{function}{}{end}

\tcp{The buffer is a priority queue}
\tcp{Storing vertices in decreasing } 
\tcp{order of buffer score}
$Q \leftarrow PriorityQueue()$

\SetAlgoNlRelativeSize{-1}

\For{$i \leftarrow 1$ \KwTo $|V|$}{
    \tcp{Reading a vertex and neighbors}
    $v , \ \N(v) \leftarrow readLine(F)$ 
    
    $v\_score \leftarrow bufferScore(v, \N(v))$

    \If{$|\N(v)| \geq D_{\text{max}}$}{
        $partitionVertex(v,    \N(v))$
    }\Else{
        $Q.push(\{v\_score, v,    \N(v)\})$
    }\If{$Q.size() == max\_qsize$}{
        
        $ t\_score, \ t , \ \N(t) \leftarrow Q.pop()$
        
        
        $partitionVertex(t,    \N(t))$
    }
}

\While{$Q.size() > 0$}{
    $t\_score , \  t,\  \N(t) \leftarrow Q.pop()$
    
    
    $partitionVertex(t,    \N(t))$
}

\function{$partitionVertex$($v, \N(v)$)}{
%
    \tcp{Finding best partition among the} \tcp{$\K$ partitions using} \tcp{partitioning score function.}
    $\V_{best} = findBestPartition(v, \N(v))$
    
    $\V_{best} = \V_{best} \cup v$
 
    $updateBufferScores(\N(v))$
}

\end{algorithm}

\textbf{Prioritized buffered streaming model.}
We take advantage of our key finding by buffering vertices that have a degree lower than a threshold, $D_{max}$. Once the buffer fills, we prioritize partitioning the vertex with the highest \textit{buffer score}.

Algorithm 1 presents the pseudocode for \textsc{Cuttana}'s prioritized buffered streaming model. The buffer, denoted by $Q$, is a priority queue sorted in descending order of buffer score and has a capacity of $max\_qsize$ vertices. The buffer score is a user-defined function designed to prevent premature assignments. Our buffer score function for $v_t$ is:

\begin{equation}
    \label{eq:bf_score}
    \frac{|\N(v_t)|}{D_{max}} + \theta  \frac{\sum_{1 \leq i \leq \K} |\N(v_t) \cap \V_i |}{|\N(v_t)|}  
\end{equation}

The rationale behind this buffer score is to assign higher buffer scores (leading to earlier eviction/placement) to vertices with more assigned neighbors, while simultaneously favoring buffering low-degree vertices. $\theta$ is a hyperparameter whose value indicates how much to favor the number of assigned neighbors over the degree. By giving more weight to the fraction of assigned neighbors, more vertices will have a chance to be buffered. However, this means the vertices will spend less time in the buffer and will be evicted with less information about their neighborhood.

When a vertex is evicted from the buffer, it needs to be assigned to a partition. This assignment can be done using the same \textit{partitioning score} function used in existing partitioners (Equation \ref{eq:score1}). For example, one could implement the streaming phase of \model using score functions from Linear Deterministic Greedy (\textsc{Ldg}) \cite{stanton2012streaming} or \fennel \cite{tsourakakis2014fennel}. In our implementation, we use the \fennel partitioning score function, with a minor adjustment to achieve more edge-balanced partitions. To select the best partition for vertex $v_t$ at time $t$, we use:

\begin{equation}
    \argmax_{1\leq i\leq \K} \left[|\V_i^t\cap\N(v_t)| - \delta \left( |\V_i^t| + \mu \sum_{x \in \V_i^t}\left|\N(x)\right| \right)\right] ,
\label{eq:scoregenfennel}    
\end{equation}

\noindent where $\delta$ is the exact penalty function used by \textsc{Fennel}. However, unlike \textsc{Fennel}, which considers only existing vertices in the partition for the penalty ($|\V_i^t|$), we adopt PowerLyra's hybrid-cut model \cite{chen2019powerlyra}, which incorporates the number of edges in the partition ($\sum_{x \in \V_i^t}\left|\N(x)\right|$) into the penalty function. Given that the number of edges exceeds the number of vertices, $\mu$ is the ratio of vertices to edges, normalizing their sum to ensure balanced growth of both vertices and edges within partitions during streaming.

After a vertex is assigned to a partition, we update the scores of its buffered neighbors, since their buffer scores have increased. We also perform a check: if all the neighbors of a vertex are assigned, we evict that vertex, a step omitted in Algorithm 1 for simplicity. The model's actual implementation includes various performance optimizations, detailed in Section \ref{sec:parallel}.


\newcommand{\s}[0]{\mathcal{S}}

\newcommand{\M}[0]{\mathcal{M}}     
\newcommand{\sub}[0]{\mathcal{S}}   
\newcommand{\PS}[0]{\mathcal{P'}}   
\newcommand{\W}[0]{\mathcal{W}}     
\newcommand{\ecp}[0]{ECP} 
\newcommand{\dec}[0]{DEC} 
\newcommand{\mvscore}[0]{MS} 
\newcommand{\move}[0]{\mathcal{MV}} 
\newcommand{\removeMVScore}[0]{remove \mvscore} 
\newcommand{\addMVScore}[0]{add \mvscore} 
\newcommand{\updECP}[0]{update \ecp} 

\subsection{Phase 2: Quality Refinement}\label{sec:p2}

\begin{definition}[Trade \& Maximality]
    We call a pair of vertex and partition index, $\langle v, b \rangle$ ($v\in V $ and $1 \leq b \leq \K$), a \textit{trade} if, after moving $v$ from its current partition to $V_b$, the total partitioning quality increases and the balance condition is maintained. If there exists no trade for a partitioning, we call the partitioning \emph{maximal}.
    
\end{definition}



\begin{figure*}[t]
\centering
\includegraphics[width=0.95\linewidth]{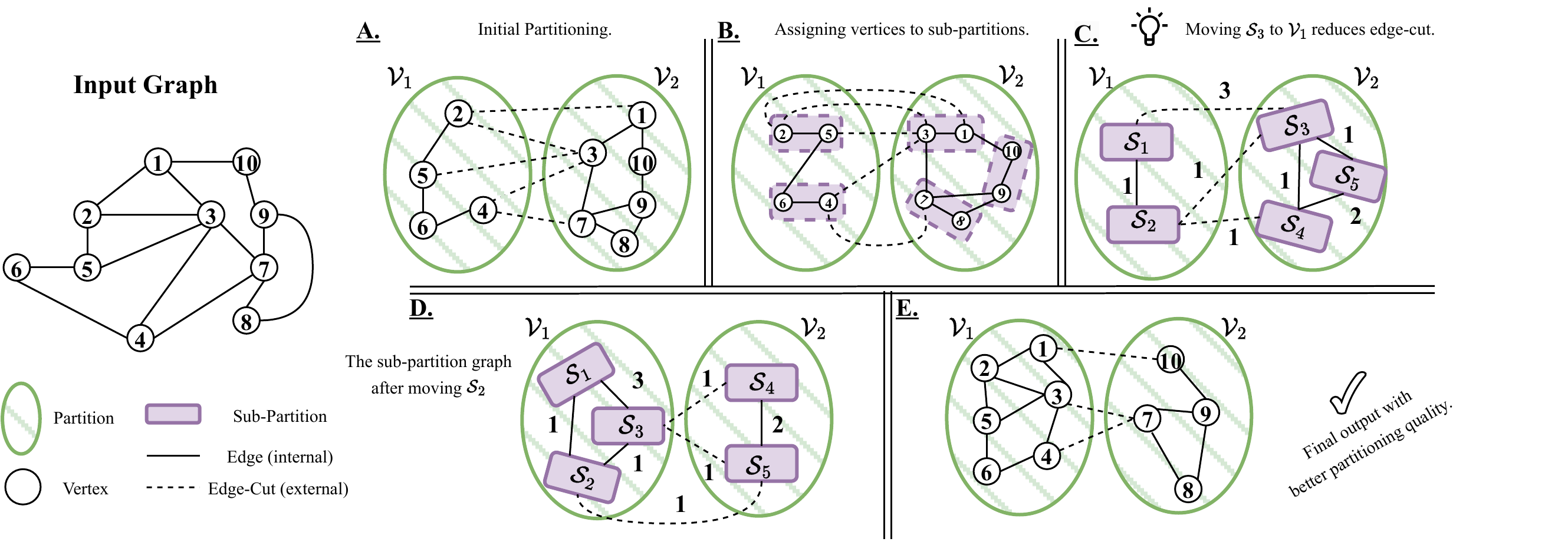}
\centering
\captionsetup{font=small} 
\caption{Partitioning of a graph and applying sub-partitioning and refinement with $\slack = 0.2$ balance condition and $\K'=5$.}
\label{fig:refine}
\end{figure*}

\begin{example}
    Figure \ref{fig:refine}.A shows the initial partitioning of a graph with balance slack $\slack = 0.2$ and the number of partitions $\K=2$. The pair $\langle 3, 1 \rangle$ is a trade since, after moving vertex 3 to $\V_1$, the total edge-cut decreases by 2 and the balance condition holds. 
\end{example}

\textbf{The quality of the streaming output is not maximal.} After partitioning a graph using a streaming partitioner, it is possible to apply trades to improve the partitioning quality, because, in practice, the balance is relaxed ($\slack > 0$), and the streaming partitioner, even with buffering, places many vertices based only on partial information. However, when applying trades, we have a more complete view of the graph. We now present our scalable refinement algorithm to enhance the partitioning produced in Phase 1 using these trades.

\textbf{Challenges of finding trades and \textit{sub-partitioning}.} Finding and applying trades requires keeping track of vertex neighborhoods. While this is possible for small graphs, it is not scalable to large graphs. To solve this problem, we coarsen the graph into a summarized version with a substantially reduced number of vertices and edges. Each coarsened vertex consists of a subset of the original vertices from the same partition. The coarsened vertices are connected with edges that are weighted according to the number of edges between their members (vertices in the original graph). We call this process \textit{sub-partitioning} and the coarsened vertices \textit{sub-partitions}.

\begin{definition}[Sub-Partitioning]
    Assuming $\K' \in \mathbb{N}$, equally-sized disjoint sets $\s_1,\s_2,\ldots,\s_{\K'}$ are a sub-partitioning of $\V_1,\V_2,$ $\ldots,\V_\K$, if $\bigcup_{1\leq i \leq \K'}\s_i = V$, and for all $\s_i$ there exists only one $\V_j$ such that $\s_i \subset \V_j$ and $\K'$ is the total number of sub-partitions.
\end{definition}

\begin{definition}[Sub-Partition Graph]
    A sub-partition graph consists of sub-partitions $\s_1,\s_2,\ldots,\s_{\K'}$ as its vertices and the edge between $\s_i,\s_j$ is a weighted edge denoted by: $$\W(\s_i,\s_j) = \left|\{\langle u,v \rangle \in E | u \in \s_i \land v \in \s_j\}\right|$$.
\end{definition}

\begin{proposition}
    The number of edge-cuts can be calculated from the sub-partition graph as the sum of $\W(\s_i,\s_j)$ for sub-partitions that are not in the same partitions. ($\left(\s_i \subset \V_{i'}\right) \land \left(\s_j \subset \V_{j'}\right)  \implies \V_{i'} \neq \V_{j'} $)
\end{proposition}

\textbf{Refinement as trades on the sub-partition graph.}
The sub-partitions can be moved between partitions via trades. Moving a sub-partition involves relocating all of its members to another partition. The goal is to reduce the total number of edge cuts, realized as a reduction of the sum of weights of the edges between sub-partitions from different partitions. We present a scalable algorithm designed to find and apply all trades in the sub-partition graph to improve the overall quality.

\textbf{Coarsening and assigning sub-partitions is another partitioning problem.} The vertices comprising a sub-partition always remain together after each trade. Our goal is to maximize the number of internal edges within a sub-partition, thereby reducing the total edges between sub-partitions. Additionally, we want to control the size of the sub-partitions and avoid skewed sizes, as such imbalances complicate maintaining the balance condition during trades. This problem mirrors the original graph partitioning problem, and we approach it similarly. We assume a constant number of subpartitions ($\frac{\K'}{\K}$) in each partition.
During Phase 1, when a vertex is placed in a partition, it is also assigned to a sub-partition within the selected partition. Any partitioning algorithm can benefit from applying refinement, which means that \textsc{Cuttana}'s Phase 1 can be implemented using any partitioning algorithm. We use the scoring function described in Equation \ref{eq:scoregenfennel} to assign vertices to sub-partitions but with different hyperparameters.

\begin{example}
    Figure \ref{fig:refine} illustrates  \textsc{Cuttana}'s refinement process. Figure \ref{fig:refine}.B is the output of phase 1 including partitioning and sub-partitioning for $\K=2, \K'=5$. Figure \ref{fig:refine}.C shows the resulting weighted sub-partition graph (coarsened graph). Figure \ref{fig:refine}.D applies the trade $\langle \s_3,1 \rangle$ on the sub-partition graph. After that, since there are no trades left, the refinement, produces the graph in Figure \ref{fig:refine}.E.
\end{example}



%
%


\textbf{Refinement Algorithm.} Although we coarsen the graph, the scalability and efficiency of the refinement algorithm determine how large we can make $\K'$. Finer-grained sub-partitions (larger $\K'$) produce better and more precise refinements. Our refinement algorithm is a greedy iterative algorithm that, in each step, applies the trade that produces the greatest quality improvement. The algorithm stops when no further trade is possible, and the partitioning is maximal. In each iteration, we consider all pairs of partitions and find the best subpartition trade among them.  To implement this algorithm efficiently, we define and use data structures that we can calculate once in Phase 1 and update efficiently during Phase 2.

Let $\PS(\s_i)$ represent the index of the partition containing $\s_i$. Let $ECP$ (edge cut per partition) be a data structure holding the number of edge cuts produced by placing a particular sub-partition in a partition. Hence, $\ecp_{\s_i,\V_{dest}}$ is the sum of all the edge weights between $\s_i$ and the sub-partitions that are not currently in partition $\V_{dest}$:
\begin{equation}
        \ecp_{\sub_i,\V_{dest}} = \sum_{1 \leq j \leq \K'} \W(\sub_i,\sub_j) \ \ [\PS(\sub_j) \neq dest]
\end{equation}



    

Next, define $\dec_{{\sub_i},{\V_{src}},{\V_{dest}}}$ as the decrease in edge-cut produced by moving $\sub_i$ from $\V_{src}$ to $\V_{dest}$, where $src = \PS(\s_i)$ and all other subpartition assignments are unchanged. When this value is negative, moving $\s_i$ to $\V_{dest}$ increases the edge-cut and worsens quality. The value of $\dec$ can be computed as:
\begin{equation}
        \dec_{{\sub_i},{\V_{src}},{\V_{dest}}} = \ecp_{\sub_i,\V_{src}} - \ecp_{\sub_i,\V_{dest}}.
\end{equation}

    

We store all $DEC$ values in the move-score structure ($MS$). Each $\mvscore_{\V_{src},\V_{dest}}$ stores all $\dec_{{\sub_i},{\V_{src}},\V_{dest}}$. To find the best trade, we iterate through all possible partition pairs $(\V_i, \V_j)$ and query $\mvscore_{\V_{i},\V_{j}}$ to determine the best trade (largest $DEC$) assuming the source partition is $\V_{i}$ and the destination is $\V_{j}$. Thus, we iterate over a total of $O(\K^2)$ move-score sets. To maintain the balance condition, we keep track of the size of each partition. If, at any move, the destination partition reaches capacity, we exclude this move from the set of possible moves.



\begin{lemma}
\label{lemma:szmvscore}
    The size of $\mvscore_{{\V_i},{\V_j}}$ and the number of sub-partitions in a partition at any point of refinement is $O(\frac{\K'}{\K})$. 
\end{lemma}
\begin{proof}
     By the definition of trade, we always maintain the balance condition, and since sub-partitions are equal-sized, a partition can have at most $(1 + \slack) \frac{\K'}{\K}$ subpartitions. The number of sub-partitions in a partition is $O(\frac{\K'}{\K})$, because $\slack$ is a small constant. Also, the size of $\mvscore_{{\V_i},{\V_j}}$ is bounded by the number of sub-partitions currently in $\V_i$.
\end{proof}

The size of each move-score set is $O(\frac{\K'}{\K})$ (Lemma \ref{lemma:szmvscore}).
We implement each move-score set as a Segment Tree \cite{de2000computational}, which means we can find 
the maximum value of a set in $O(1)$ and update it in 
$O(log(\frac{\K'}{\K}))$.
Updating is implemented by deleting the $\dec$ value and inserting a new value.

\textbf{Updating Variables After a Trade.} The main challenge in the refinement algorithm is efficiently updating $MS$.
Moving $S_x$ from $\V_{src}$ to $\V_{dest}$ involves updating the $\ecp$ values and changing the $\dec$ values stored in $\mvscore$. We need to update $\ecp$ only for the vertices that are neighbors of $\s_x$. For any neighbor $\s_i \in \N(S_x)$, we perform:

\begin{equation}
    \begin{aligned}
& \ecp_{\sub_i,\V_{src}} = \ecp_{\sub_i,\V_{src}} + \W(\sub_i,\sub_x) \\
& \ecp_{\sub_i,\V_{dest}} = \ecp_{\sub_i,\V_{dest}} - \W(\sub_i,\sub_x) \\
\end{aligned}
\end{equation}
In the worst case, $\s_x$ can be neighbors to all other sub-partitions, making this step $O(\K')$. 

Updating $\dec_{{\s_i},{\p'(\s_i)},{\V_j}}$ naively can result in $O(\K'\K)$ updates. Worse yet, changing each entry in the move-score set is $O(log(\frac{\K'}{\K})$, so the naive approach is $O(\K'\K \log(\frac{\K'}{\K}))$ in total, because the moved sub-partition can have $O(\K')$ neighbors, and those neighbors can go from their partitions to $O(\K)$ other partitions. However, it can be done in $O(\K' \log(\frac{\K'}{\K}))$ by exploiting the following theorem.

\begin{theorem}
\label{teo:cost}
After applying each trade, we require updating only $O(\K')$ entries in the move-score sets.
\end{theorem}
\begin{proof}
    After moving $\s_x$, we categorize its neighbours $\s_i \in \N(\s_x)$ into two cases:
    \begin{enumerate}
        \item $\p'(\s_i) \in \{{src}, {dest}\}$: In this case, for all of the partitions $\V_j$, we have to update $\dec_{{\s_i},{\p'(\s_i)},{\V_j}}$ . However, because we have $O(\frac{\K'}{\K})$ sub-partitions in both $\V_{src},\V_{dest}$ (Lemma \ref{lemma:szmvscore}), updating $\dec_{{\s_i},{\p'(\s_i)},{\V_j}}$, even for all of the partitions $\V_j$, is $O(\K.\frac{\K'}{\K})$ or $O(\K')$.
        \item $\p'(\s_i) \notin \{{src}, {dest}\}$: In this case, if the neighbor is in neither the source nor destination partitions, we need to only update $\dec_{{\s_i},{\p'(\s_i)},{\V_j}}$ where $\V_j \in \{\V_{src}, \V_{dest}\}$, because the number of edges from $\s_i$ to the sub-partitions in other partitions are unchanged. Since the number of neighbors is bounded by the number of sub-partitions, we also have to perform $O(\K')$ updates, but this time only for two target partitions. 
    \end{enumerate}

\end{proof}

Finally, we have to update $\dec$ variables and move-score sets for $\s_x$ itself. For all partitions $\V_j$, we have to remove all of $\dec_{\s_x,\V_{src},\V_j}$ from $\mvscore_{\V_{src}, \V_j}$ and add all $\dec_{\s_x,\V_{dest},\V_j}$ to $\mvscore_{\V_{dest}, \V_j}$,
since $\s_x$ changed its partition.
In summary, we find the best move for each iteration in $O(\K^2)$ and update all variables in $O(\K'\log(\frac{\K'}{\K}))$. The number of refinement steps is finite, as we decrease edge cut in each step, and edge-cut is finite. The algorithm stops at the maximal partitioning when there is no move left. However, due to the coarse granularity of sub-partitions and weighted edges, in practice, the improvements are also coarse-grained. It is possible to stop refinement process early when the best move improves edge-cut by less than a threshold ($Thresh$). This early stop provides a time/quality trade-off, and the threshold produces a worst-case bound for the number of steps of $\frac{|E|}{Thresh}$, since the upper bound for edge-cuts is $|E|$, and each step improves at least $Thresh$ edge-cuts.



\begin{figure}[t]
\centering
\includegraphics[width=0.90\linewidth]{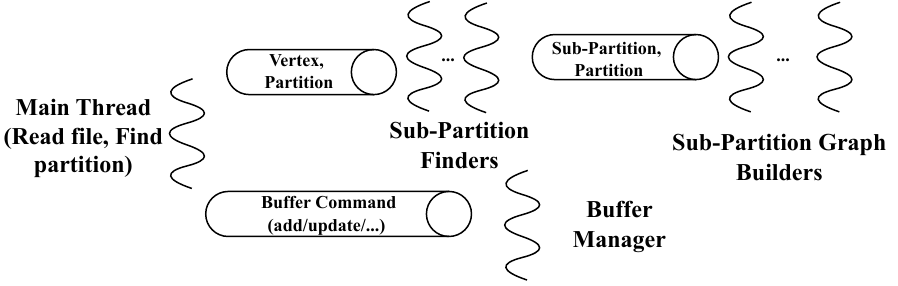}
\centering
\captionsetup{font=small} 
\caption{Parallel model of \model execution.}
\label{fig:parallel}
\end{figure}

\subsection{Parallel Partitioning and Implementation}\label{sec:parallel}
\textbf{Parallelization.} Existing streaming partitioners use only a single thread. This underutilizes resources in modern multicore computers. We leverage the unused cores to parallelize \model, thereby reducing the overhead introduced by buffering and refinement. Our approach involves dividing the computational load in such a way that different threads do not write to shared variables, thus avoiding the need for locking. Thread communication is facilitated using hardware-optimized lock-free queues \cite{valois1994implementing,lfqueue}. At a high level, Phase 1 runs in parallel, partitioning, sub-partitioning, and building the data structures for refinement; Phase 2 simply applies trades.

The primary thread reads the file, selects vertices for buffering, and partitions them after eviction from the buffer. Once a vertex is assigned to a partition, other threads are notified to determine the sub-partition of that vertex and update variables for refinement (Figure \ref{fig:parallel}). Additionally, we have a buffer manager thread that pushes/pops the buffer based on main-thread commands and applies changes to the buffer score whenever a new vertex has been partitioned. Threads of the same kind (i.e., subpartition finders) shard the shared variables based on a key (e.g., vertex ids, sub-partition id). 

\textbf{Implementation.} The \textsc{Cuttana} software package is implemented in approximately 1500 lines of C++. Users can specify the number of sub-partitions and buffer size, allowing for a quality/time trade-off. Furthermore, \model offers two modes: edge-balance and vertex-balance. In edge-balance mode, capacities and sizes are calculated based on the number of edges in the partition. Due to space constraints, we have omitted details regarding minor implementation-level optimizations for quality and partitioning latency improvement (e.g., applying parallel changes for updating data structures and variables).







\section{Experimental Analysis}
Our experimental analysis answers the following research questions.
\begin{itemize}
    \item \textbf{RQ1:} How does \model partition quality compare to that of existing approaches?
   \item \textbf{RQ2:} Given that existing vertex partitioners impose a vertex-balance constraint, how much edge imbalance do they produce? How does the partition quality change if they adopt an edge-balance constraint? 
    \item \textbf{RQ3:} How much do buffering and refinement affect partition quality?
    \item \textbf{RQ4:} How does \model partitioning affect the performance of Distributed Graph Analytics?
        \item \textbf{RQ5:} How does \model partitioning affect the performance of a Distributed Graph Database?
\end{itemize}

\textbf{Datasets.}
Table \ref{table:dataset} shows the characteristics of datasets used in our study. We selected graphs of different sizes and domains to represent various use cases. The web graphs are hyperlink networks where vertices are webpages and the edges are links. In social networks, the vertices are users and the edges are follow/friend relationships. All of the datasets were obtained from the Konect network repository \cite{konect} except for the LDBC social network benchmark, which we obtained using the LDBC generator \cite{erling2015ldbc}, the US-Roads dataset \cite{usroad}, and the RMAT synthetic dataset which we generated using ParMAT \cite{wsvr}. We used real-world natural graphs, including both big and small graphs, to analyze the quality of partitioning for different algorithms. We utilized large graphs, both real-world and synthetic, for distributed graph analytics, and finally, we used the LDBC benchmark for our graph database evaluation.


\renewcommand{\arraystretch}{1.2}
\setlength{\arrayrulewidth}{0.2mm}
\begin{table}[]
\centering
\caption{Graph datasets used in the evaluation.}
\begin{tabular}{|l||r|r|c|}

\hline
\textbf{Name}                                            & \# \textbf{Vertices}                         & \# \textbf{Edges}                              & \textbf{Domain}         \\ \hline \hline
US-Roads (\textbf{usroad})                                    & 23M  & 28M                    & Road    \\
Orkut (\textbf{orkut})                                      & 3M   & 117M   & Social  \\
UK domains - 2002 (\textbf{uk02})                        & 18M  & 261M   & Web       \\ 
LDBC-SNB-SF1000 (\textbf{ldbc})                          & 3M   & 490M                           & Social  \\
RMAT Large (\textbf{rMat-XL})) & 10M & 1B & Synthetic \\
Twitter (\textbf{twitter})                                    & 41M  & 1.4B & Social  \\

{UK domains - 2007 (\textbf{uk07})} & 105M & 3.3B & Web       \\ \hline
\end{tabular}
\label{table:dataset}
\end{table}

\begin{landscape}

\renewcommand{\arraystretch}{1.3}
\setlength{\arrayrulewidth}{0.3mm}
\begin{table*}[t]

\centering
\captionsetup{justification=centering} 
\caption{Analysis of Partitioning Quality on eight partitions ($\K = 8$). The boldfaced
numbers shaded blue indicate the best result for each graph and balance condition. The Improv. column shows the improvement of \model over \textsc{Fennel}.}
\begin{tabular}{c|l|cccc||cccc||cc}
\hline
\multirow{2}{*}{\shortstack{Quality \\ Metric}} & \multirow{2}{*}{Dataset} & \multicolumn{4}{c||}{Edge-Balance Condition (\textit{EB}) ($\slack = 0.10$)}                & \multicolumn{4}{c||}{Vertex-Balance Condition (\textit{VB}) ($\slack = 0.05$)}                     & \multicolumn{2}{c}{Improv. }         \\
                        &                          & \model             & \textsc{Fennel} & \textsc{HeiStream}      & \textsc{\textsc{Ldg}}   & \model              & \textsc{Fennel} & \textsc{HeiStream}            & \textsc{\textsc{Ldg}}   & \textit{EB} & \textit{VB}\\ \hline
\multirow{5}{*}{\shortstack{\textit{edge-cut} \\ $\lambda_{EC}$ \\ (\%)}}     & \textit{usroad}                   & 27.93               & 31.22  & \cellcolor{blue!15}\textbf{16.84} & 30.06 & 22.5                 & 31.15  & \cellcolor{blue!15}\textbf{10.48} & 30.05 & \multicolumn{1}{c|}{11\%} & 28\%             \\
                        & \textit{orkut}                    & \cellcolor{blue!15}\textbf{39.3}       & 50.33  & 55.22          & 57.43 & { \cellcolor{blue!15}\textbf{32.33}} & 43.31  & 42.15                & 53.11 & \multicolumn{1}{c|}{22\%} & 26\%            \\
                        & \textit{uk02}                     & { \cellcolor{blue!15}\textbf{3.03}} & 3.91   & 17.7           & 14.53 & \cellcolor{blue!15}\textbf{3.26}        & 7.12   & 10.05                & 16.3  & \multicolumn{1}{c|}{23\%} & 55\%             \\
                        & \textit{twitter}                  & \cellcolor{blue!15}\textbf{64.21}      & 68.39  & 64.67          & 73.04 & { \cellcolor{blue!15}\textbf{34.09}} & 37.80  & 45.62                & 55.9  & \multicolumn{1}{c|}{6\%}  & 10\%             \\
                        & \textit{uk07}                     & \cellcolor{blue!15}\textbf{1.64}       & 2.73   & 21.9           & 11.71 & { \cellcolor{blue!15}\textbf{1.4}}   & 3.35   & 6.65                 & 12.11 & \multicolumn{1}{c|}{40\%} & 59\%             \\ \hline \hline
\multirow{5}{*}{\shortstack{\textit{communication} \\ \textit{volume} \\ $\lambda_{CV}$ \\ (\%)}}     & \textit{usroad}                   & 7.93                & 9.06   & \cellcolor{blue!15}\textbf{4.41}  & 8.68  & 6.09                 & 9.04   & \cellcolor{blue!15}\textbf{2.97}  & 8.68  & \multicolumn{1}{c|}{13\%} & 33\%             \\
                        & \textit{orkut}                    & \cellcolor{blue!15}\textbf{44.82}      & 63.83  & 65.48          & 63.42 & { \cellcolor{blue!15}\textbf{44.09}} & 55.95  & 48.43                & 61.01 & \multicolumn{1}{c|}{30\%} & 22\%             \\
                        & \textit{uk02}                     & \cellcolor{blue!15}\textbf{4.25}       & 5.45   & 6.74           & 6.74  & 4.68                 & 5.63   & { \cellcolor{blue!15}\textbf{3.78}}  & 6.78  & \multicolumn{1}{c|}{22\%} & 17\%             \\
                        & \textit{twitter}                  & \cellcolor{blue!15}\textbf{40.91}      & 43.72  & 50.23          & 46.77 & \cellcolor{blue!15}\textbf{41.3}        & 47.04 &  44.04                  & 47.39 & \multicolumn{1}{c|}{6\%}  & 13\%              \\
                        & \textit{uk07}                     & \cellcolor{blue!15}\textbf{4.5}        & 7.21   & 8.98           & 6.12  & { \cellcolor{blue!15}\textbf{3.88}}  & 5.29   & 3.99                 & 6.01  & \multicolumn{1}{c|}{38\%} & 27\%             \\ \hline
\end{tabular}
\label{table:contrib}
\end{table*}

\end{landscape}

\textbf{Baselines.}
We compare \model to three other streamining partitioners: \fennel \cite{tsourakakis2014fennel}, \ldg \cite{stanton2012streaming}, and \heistream \cite{faraj2022buffered}. \fennel and \ldg are score-based streaming vertex partitioners. \fennel is the best baseline to show the benefits of buffering and refinement, since \model is implemented on top of \fennel and uses the same scoring function. \heistream is recent work that has the same motivation as \textsc{Cuttana}, i.e., bridging the gap between streaming and in-memory solutions in quality and scalability. \heistream reads and assigns vertices in batches and claims to beat \fennel \cite{faraj2022buffered}. We used the implementation of \fennel and \ldg provided by Pacaci \cite{pacaci2019experimental} since the official code is not available, and we obtained \heistream from the authors.  In graph analytic benchmarks, where it's possible to use edge-partitioners \cite{pacaci2019experimental}, we also compare \model to \textsc{Hdrf} \cite{petroni2015hdrf} and \textsc{Ginger} \cite{chen2019powerlyra}.

\begin{figure*}[ht]
  \centering
  \begin{subfigure}{\textwidth}
    \includegraphics[width=0.98\linewidth, height=0.75\textheight, keepaspectratio]{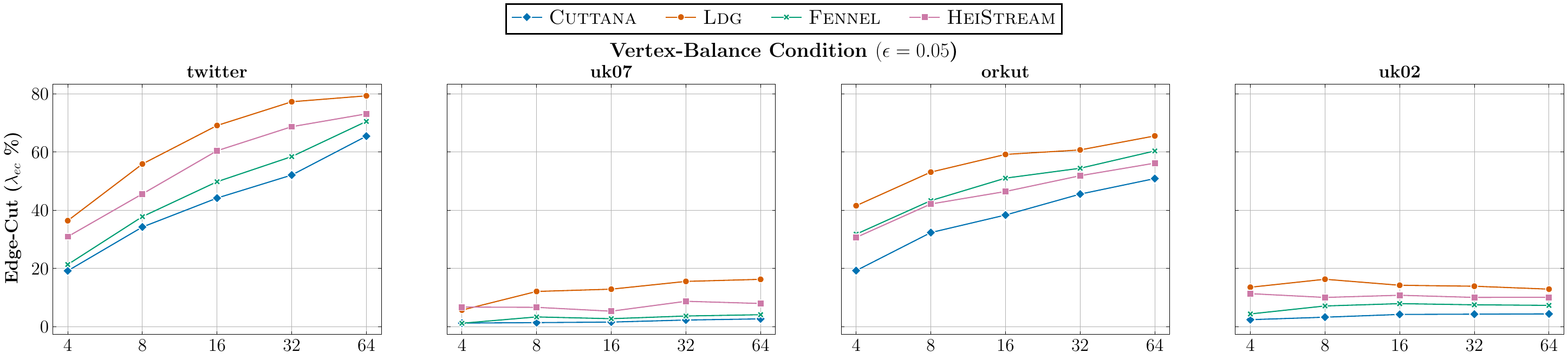}
    \label{fig:sub1}
  \end{subfigure}


  \begin{subfigure}{\textwidth}
    \includegraphics[width=0.98\linewidth, height=0.8\textheight, keepaspectratio]{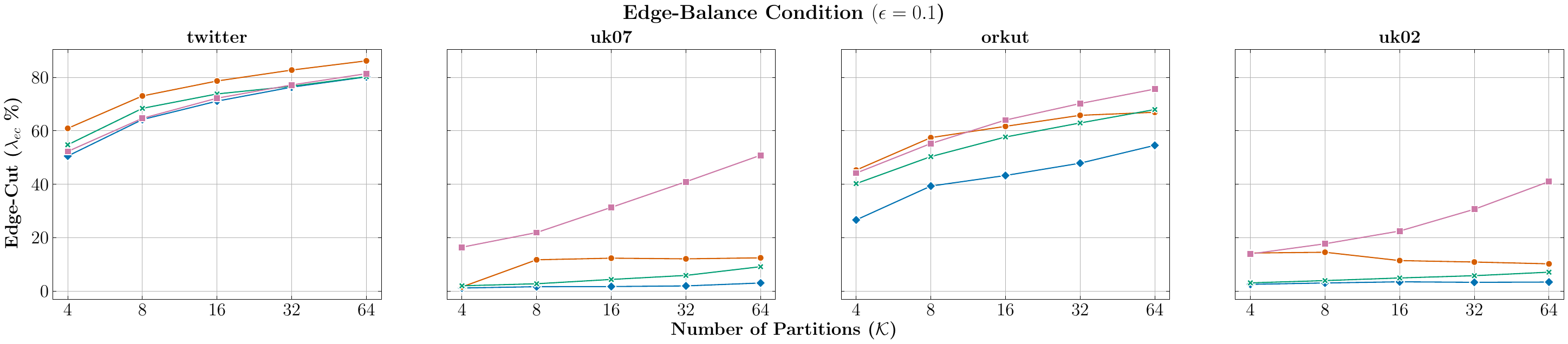}
    \label{fig:sub2}
  \end{subfigure}

  \caption{The partitioning quality of \model excels consistently across varying numbers of partitions.}
  \label{fig:perk}
\end{figure*}

\textbf{Experimental Setup and Reproducibility.} 
We conducted all our experiments, including partitioning, analytics, and graph database benchmarks, on a private cluster of 16 machines, each equipped with an 8-core \text{Intel\textsuperscript{\textregistered} Xeon\textsuperscript{\textregistered} Silver 4309Y Processor} and 64 GB of RAM. Unless otherwise specified, we ran \model with $D_{\text{max}}=1000$, $max\_qsize=10^6$ vertices (consuming at most 4 GB of DRAM) and determined the number of sub-partitions such that $\frac{\K'}{\K} = 4096$ for all datasets except Twitter on which we 
set $D_{\text{max}}=100$ and determined the number of sub-partitions such that $\frac{\K'}{\K} = 256$. 
The code for \model and the framework for the application study (analytics, databases) are publicly available. We conducted the application study based on the benchmarking framework provided by Pacaci and Ozsu \cite{pacaci2019experimental}, making minor modifications to update deprecated packages, add support for \heistream and \textsc{Cuttana}, and add some additional features. The partitioning process is deterministic as we fixed the random seed used for tie-breaking among partitions with the same score. We also disabled the buffer-manager thread. The use of a buffer-management thread introduces scheduling randomness, which we disable by offloading the task to the main thread to ensure reproducibility. We used the baselines with default hyperparameters.


\subsection{Quality Metrics Analysis}

\textbf{Improving Edge-cut and Communication Volume.} We address \textbf{RQ1} by partitioning datasets under both edge/vertex-balance constraints using all baseline algorithms. We measure the communication volume and edge-cut as indicators of network overhead in distributed applications with/without message aggregation, respectively. Table \ref{table:contrib}
shows that \model produces better quality partitions in nearly all scenarios. The benefit is most pronounced for the largest graphs (twitter and uk07) in edge-balance mode, suggesting that it is possible to effectively partition massive graphs that cannot be partitioned by in-memory partitioners.
Since the reported metrics are normalized, their relative difference $\left(\frac{|\lambda_1 - \lambda_2|}{\max(\lambda_1,\lambda_2)}\right)$ is an underestimate of the reduction in network overhead.

\textsc{Cuttana} consistently improves partitioning quality from 6\% to 59\%. This improvement reflects network overhead, which is the dominant overhead in distributed graph processing, so we anticipate a more significant improvement in end-to-end application latency as well. In large graphs such as Twitter and UK07, \model produces better partitioning quality than \heistream by up to 19\%  and 93\%, respectively. However, in the US-Roads datasets, \textsc{HeiStream} produces better partition quality than \textsc{Cuttana}. \textsc{Heistream}'s authors told us that the algorithm is sensitive to ordering and performs best when each batch consists of vertices from the same neighborhood with many edges among them. The size and original ordering of US-Roads are ideal for \textsc{HeiStream}. On the other hand, \textsc{Cuttana}'s buffering is robust to input order; the only case in which \textsc{Cuttana} does not provide the lowest edge-cut (communication volume) is when the original input order happens to be ideal for \textsc{HeiStream}.

Figure \ref{fig:perk} shows partition quality as a function of the number of partitions.
While \fennel and \heistream outperform each other depending on the dataset and balance condition, \model outperforms both.

\textbf{The Case for Edge-Balance using Vertex Partitioners.} Pacaci and Ozsu \cite{pacaci2019experimental} uncovered two key properties of state-of-the-art vertex partitioners. They demonstrated that both \fennel and \ldg exhibit lower network overhead than edge partitioners, but they suffer from significant worker imbalance, rendering them less compelling. In some scenarios, random partitioning produced better application performance due to its superior load balancing. We determined that the root cause lies in using a vertex-oriented balance constraint. Balancing the number of vertices in a partition does not necessarily balance the number of edges.
However, edge-balance is crucial from a computational load-balancing perspective, because almost all graph algorithms iterate over edges.
In other words, the number of edges in a partition determines the workload on each participant in a distributed computation, and edge imbalance leads to stragglers.
Edge balance is more critical than vertex balance since the number of edges dominates the number of vertices.
Moreover, the balanced assignment of edges is crucial in memory-constrained scenarios.
When the number of edges in each partition varies, we must either over-provision memory (which is expensive) or suffer the consequences that some workers will be computing on out-of-memory data, thus exacerbating the delay that stragglers impose.

\begin{figure}[t]
\centering
\includegraphics[width=0.65\linewidth]{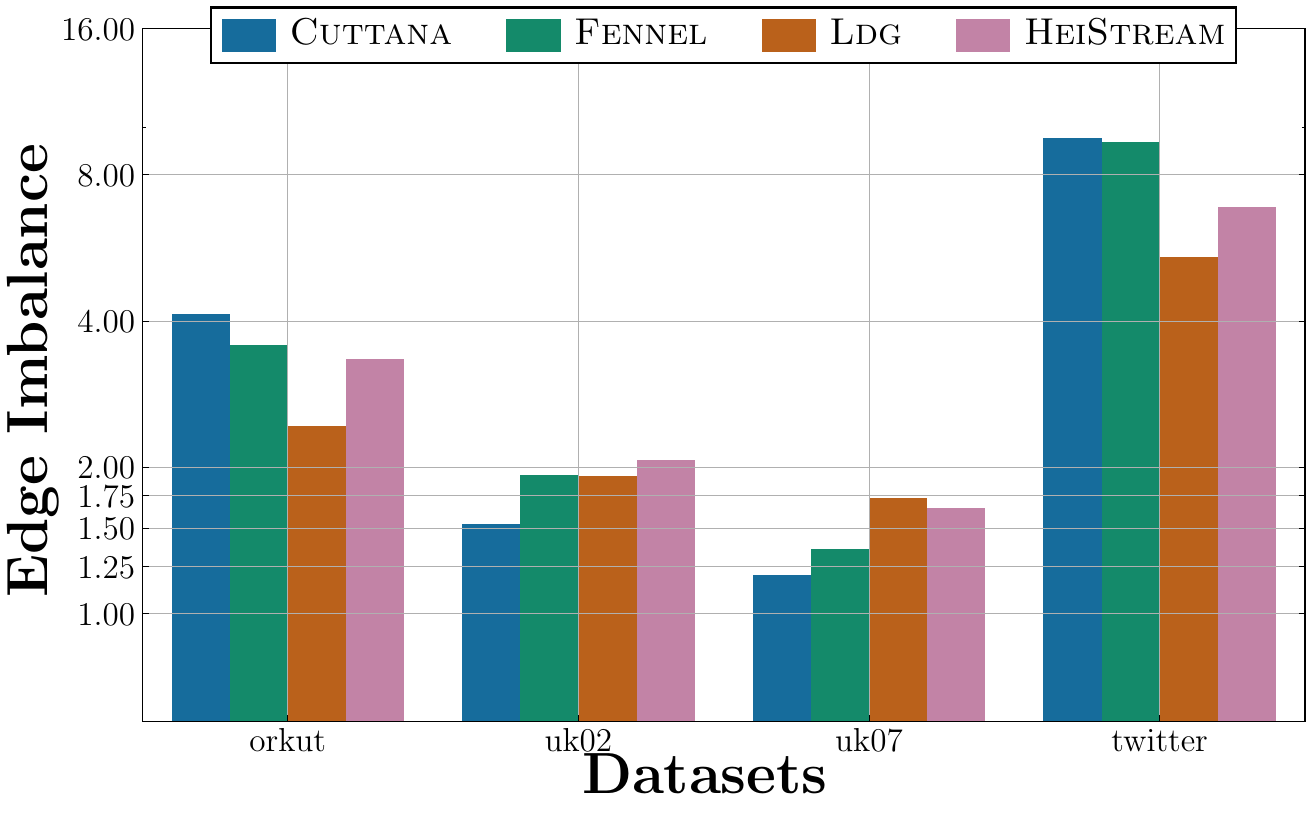}
\centering
\captionsetup{justification=centering,font=small} 
\caption{Baselines partitioners and \model when using a vertex-balance condition (which is not \model's default) can lead to uncontrolled edge-imbalance and uneven load distribution.}
\label{fig:ebshow}
\end{figure}

We demonstrate this issue in Figure \ref{fig:ebshow}. We use all the baselines and modify \model to use a vertex-balance constraint instead of its preferred edge-balance constraint.
We set $\slack = 0.05$ and show the ratio of the maximum number of edges in any partition
to the average number of edges across all partitions.
Although the vertices are balanced among the partitions, the edges are hugely imbalanced. This suggests that, regardless of the partitioning scheme and dataset, using the vertex-balance condition, which is prevalent in the literature\cite{tsourakakis2014fennel,stanton2012streaming,abbas2018streaming,awadelkarim2020prioritized,faraj2022buffered}, yields partitions with too many edges, leading to stragglers when computing in parallel. For example, on Twitter, using all of the partitioners in vertex balance mode causes one worker machine to have at least \textit{4x} more load than other machines. Overweight partitions also risk producing more network overhead. \model offers the user both vertex and edge balance options. \heistream was originally implemented for vertex-balance, but the authors added the edge-balance feature upon our request. We added edge-balance support to \fennel and \ldg using the same approach as that used in \textsc{Cuttana}.

The answer to \textbf{RQ2} can also be found in Figures \ref{fig:perk} and \ref{fig:ebshow} and Table \ref{table:contrib}, which show that 1) satisfying edge balance makes edge cut worse, and 2) vertex-balance produces significant edge imbalance (as discussed above). 
However, \model produces the best partition quality when satisfying either balance constraint.
In the rest of our evaluation, we use \model's edge-balance mode and the original baseline implementations. Experimentally, especially for communication volume and on graphs other than Twitter, the additional overhead of quality difference introduced by satisfying edge-balance was amortized in execution time by having more even computation and network overhead distribution.


\begin{table}
\label{table:abl}
\centering
\caption{Contribution of different components of \model to the final partitioning quality ($\K = 16$). The numbers represent the normalized edge-cut ($\lambda_{EC}$), and the percentages indicate the improvement over \fennel (i.e., \model without refinement and buffering).}
\begin{tabular}{|l|cccc|}
\hline
Algorithm                                                                             & Orkut                                                  & Twitter                                                & UK07                                                  & UK02                                                  \\ \hline \hline

\model                                                                              & \begin{tabular}[c]{@{}c@{}}38.3 \\ (25\%)\end{tabular} & \begin{tabular}[c]{@{}c@{}}44.1 \\ (11\%)\end{tabular} & \begin{tabular}[c]{@{}c@{}}1.5 \\ (52\%)\end{tabular} & \begin{tabular}[c]{@{}c@{}}2.7 \\ (66\%)\end{tabular} \\ 
\hline

\model w/o Refine                                                                    & \begin{tabular}[c]{@{}c@{}}40.7 \\ (20\%)\end{tabular} & \begin{tabular}[c]{@{}c@{}}47 \\ (6\%)\end{tabular}    & \begin{tabular}[c]{@{}c@{}}1.7 \\ (45\%)\end{tabular} & \begin{tabular}[c]{@{}c@{}}4.9\\  (38\%)\end{tabular} \\ \hline
\model w/o Buffer                                                                    & \begin{tabular}[c]{@{}c@{}}45.9 \\ (10\%)\end{tabular} & \begin{tabular}[c]{@{}c@{}}48.2 \\ (3\%)\end{tabular}  & \begin{tabular}[c]{@{}c@{}}2\\  (35\%)\end{tabular}   & \begin{tabular}[c]{@{}c@{}}6.2 \\ (22\%)\end{tabular} \\ 

\hline

\begin{tabular}[c]{@{}c@{}}\model w/o Buffer \& \\ Refinement  (\textsc{Fennel})\end{tabular} & 51                                                     & 49.8                                                   & 3.1                                                   & 7.9                                                   \\ \hline
\end{tabular}

\end{table}

\textbf{Ablation Study \& Partitioning Latency.} We analyzed the isolated contributions of the two main components of \textsc{Cuttana}, as shown in Table 3, to answer \textbf{RQ3}. Generally, buffering was the main contributor to quality improvement. The relative improvement of refinement was higher when there was no buffering and the initial partitioning had lower quality.

\begin{figure}
  \centering

    \includegraphics[width=\linewidth,height=0.4\linewidth]{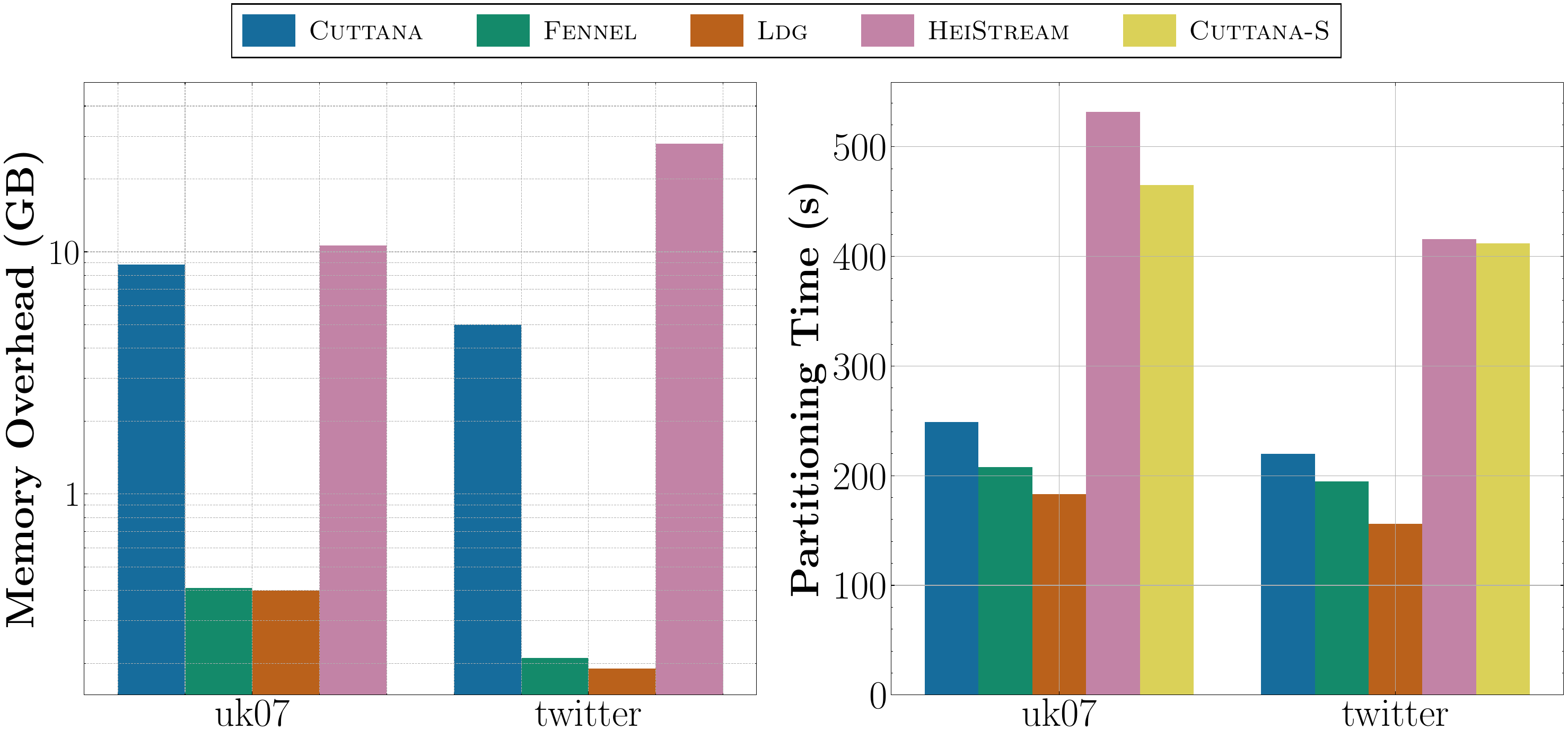}

\captionsetup{justification=centering,font=small,skip=0pt} 
  \caption{Memory overhead (log-scale) and time efficiency of \textsc{Cuttana} compared to baselines and single-thread implementation (\textsc{Cuttana-S})}
  \label{fig:timeeff}
\end{figure}

Figure \ref{fig:timeeff} compares \textsc{Cuttana}'s memory consumption and partitioning time to the baselines. The memory overhead is high relative to \fennel and \textsc{Ldg}; however, this is not a cause for concern as the overhead is independent of graph size and consumes only a small fraction of the main memory available on today's commodity computers. \model has a small additional time overhead compared to \fennel and is nearly twice as fast as \heistream for large graphs. In Table 4, we demonstrate that we more than compensate for \textsc{Cuttana}'s time overhead, relative to \fennel, by running analytic tasks much more quickly.

\begin{figure}[t]
\captionsetup{justification=centering,font=small} 
  \begin{subfigure}{0.47\textwidth}
    \centering
    \includegraphics[width=\textwidth]{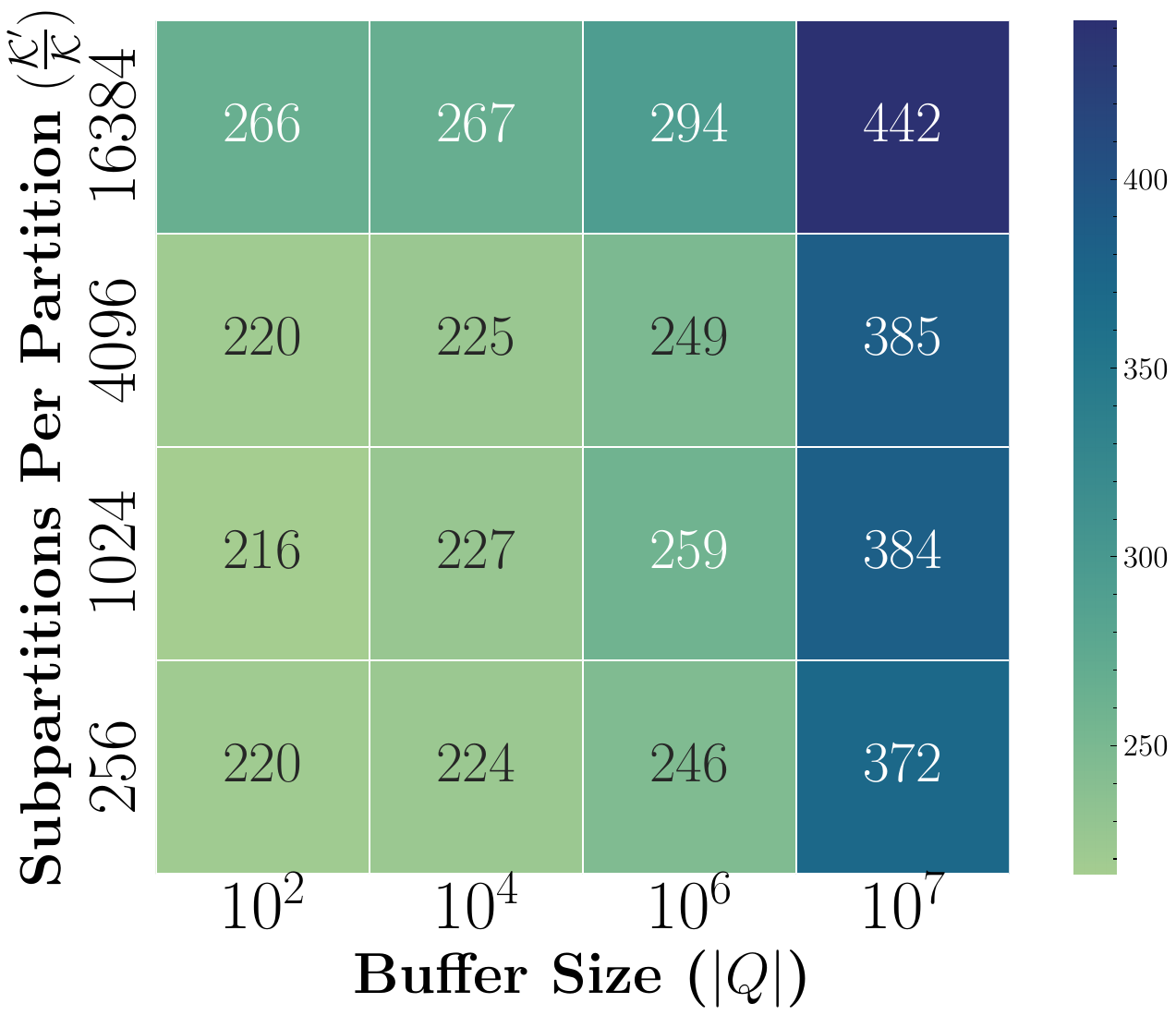}
    \label{subfig:3}
  \end{subfigure}
  \hfill
  \begin{subfigure}{0.47\textwidth}
    \centering
    \includegraphics[width=\textwidth]{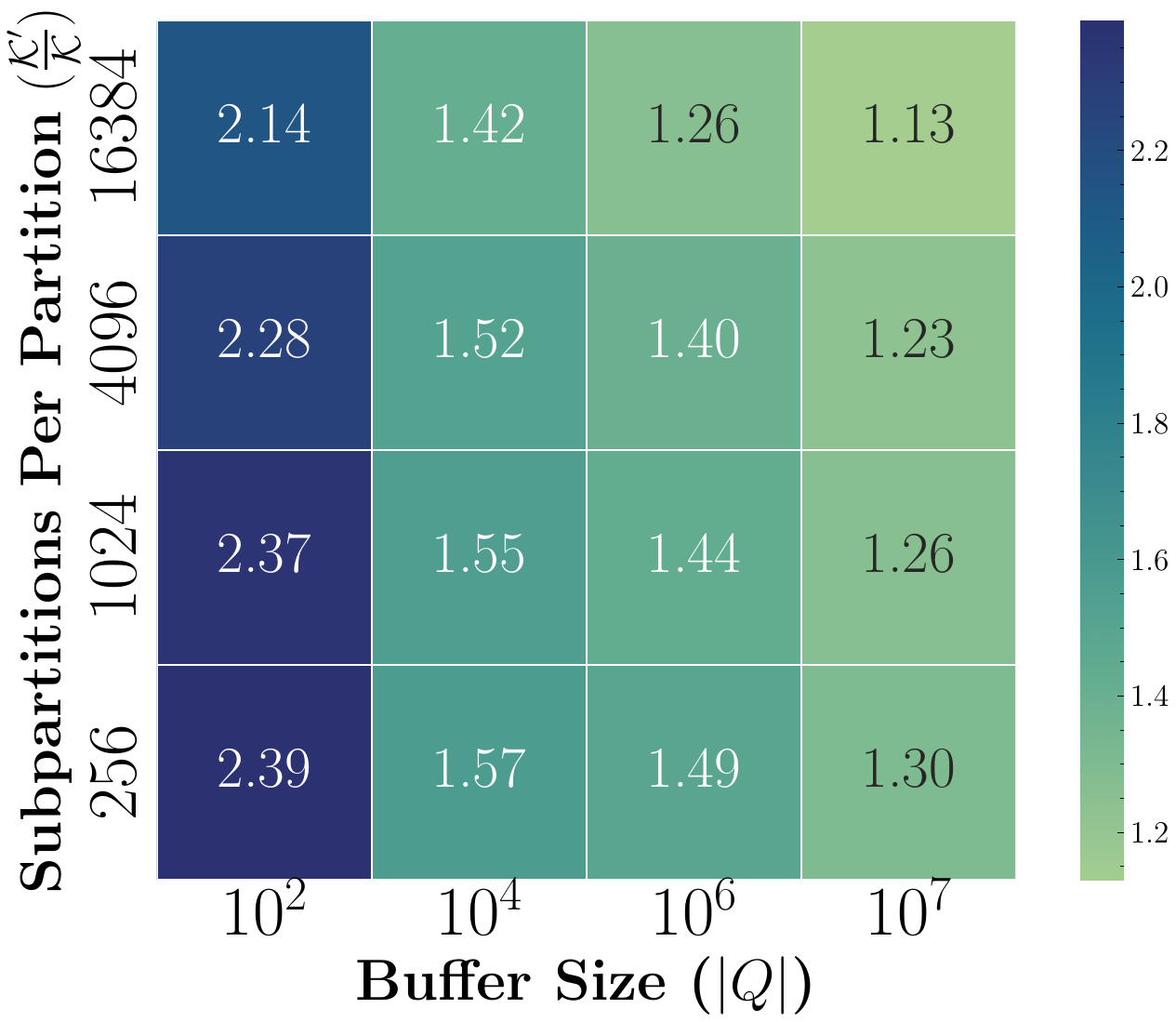}
    \label{subfig:4}
  \end{subfigure}
\captionsetup{justification=centering,font=small,skip=0pt} 

  \caption{Impact of Buffer Size on Partitioning Quality and Time for uk07}
  \label{fig:heatmap}
\end{figure}

Figure \ref{fig:heatmap} highlights the tradeoffs that a user can make when configuring \textsc{Cuttana}.
Both partitioning time and memory consumption are governed by the selection of the buffer size, $|Q|$, and the number of subpartitions, $K'$.
\model performs more work than \fennel due to buffering, updating buffer scores, selecting sub-partitions, and updating the data structures we use to optimize refinement. Using lock-free queues, we assign these tasks to a background thread. This reduces the overhead of enqueuing requests, draining the buffer at the end of streaming, and applying refinement.

\begin{figure}[t]
\captionsetup{justification=centering,font=small} 
  \begin{subfigure}{0.47\textwidth}
    \centering
    \includegraphics[width=\textwidth]{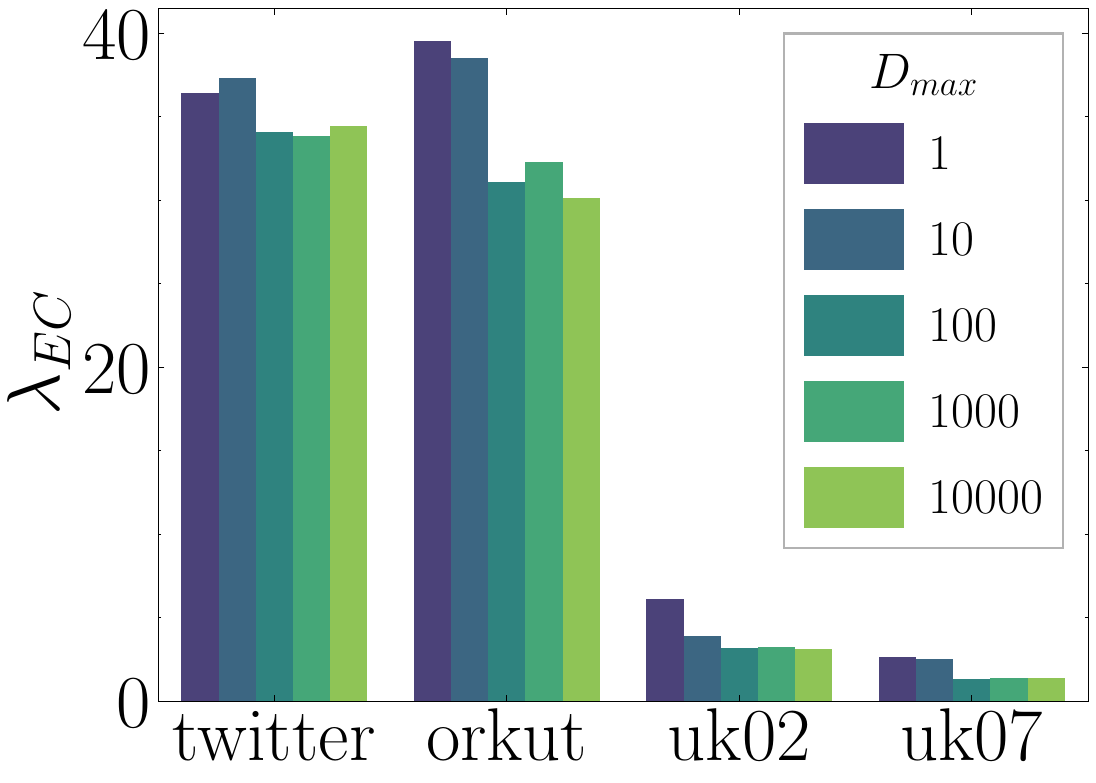}
    \label{subfig:3}
  \end{subfigure}
  \hfill
  \begin{subfigure}{0.47\textwidth}
    \centering
    \includegraphics[width=\textwidth]{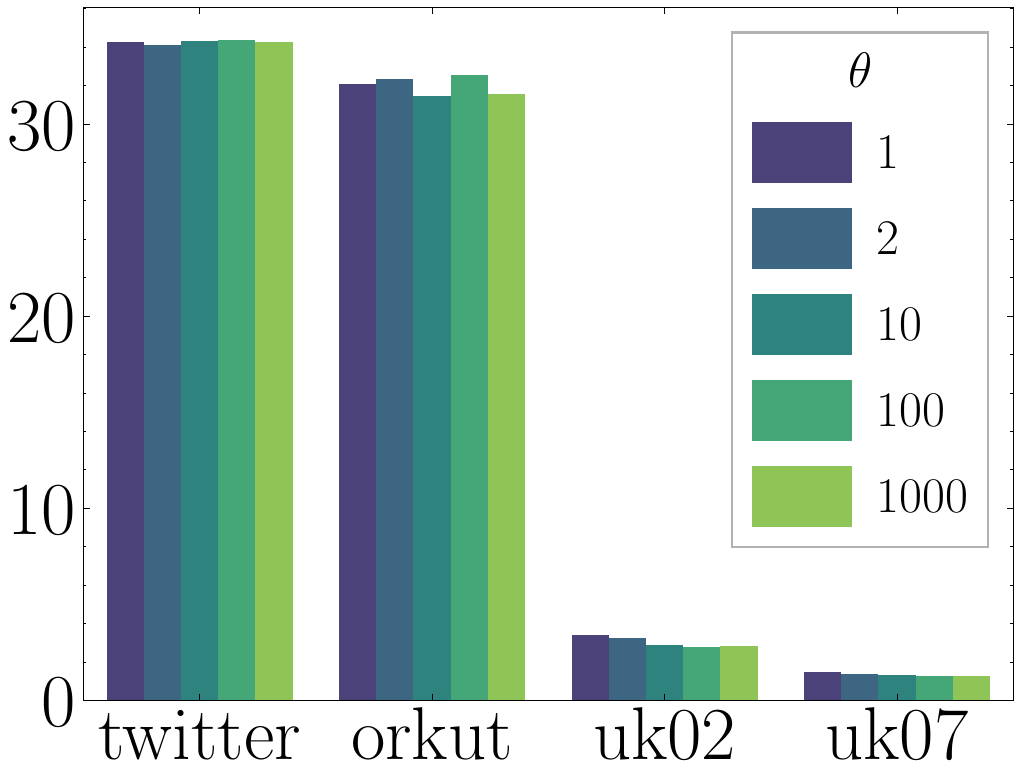}
    \label{subfig:4}
  \end{subfigure}
\captionsetup{justification=centering,font=small,skip=0pt} 

  \caption{Effect of $D_{max}$ and $\theta$ on Partitioning Quality}
  \label{fig:hyperparam}
\end{figure}

Large buffer sizes and numbers of sub-partitions can cause the background thread to fall behind the main thread.
If the queues fill, then the main thread blocks. \model maintains competitive performance up to 4096 sub-partitions and $10^6$ vertices in the buffer.
We use these as the default values. While selecting larger parameters can yield higher quality partitions, it should be undertaken cognizant of the impact on partitioning time.
Figure \ref{fig:hyperparam} shows the effects of the buffering threshold, $D_{max}$, 
and the score function scale parameter, $\theta$.
$D_{max}$ should be set to a value between 100 and 1000 to produce both good quality and performance.
As shown in Figure \ref{fig:powerlaw}, this lets us store the majority of vertices in the buffer, due to the cumulative distribution of power-law graphs. A lower threshold makes it impossible for many vertices to avoid premature assignment, significantly affecting partitioning quality. Higher thresholds can provide minor gains in partitioning quality, but they negatively impact performance and memory usage. $\theta$ has a minor impact on partition quality. Our general advice is to use $100 \leq D_{max} \leq 1000$ and $2 \leq \theta \leq 10$.

\subsection{Application Study}

\begin{landscape}
\begin{table*}[ht]
\centering
\caption{The latency, in seconds, of PageRank (PR), Connected Components (CC), and Single Source Shortest Path (SSSP) workloads using different partitioners. The boldfaced numbers shaded blue indicate the best result for each workload.}
\label{table:analytic}
\begin{tabular}{|cc||cccccc||c|c|}
\hline
\multicolumn{2}{|c|}{\multirow{2}{*}{\begin{tabular}[c]{@{}c@{}}Dataset/\\ Algorithm\end{tabular}}} & \multicolumn{6}{c||}{Partitioning Scheme}                                     & \multirow{2}{*}{\begin{tabular}[c]{@{}c@{}}Performance over the \\ best Vertex Partitioner\end{tabular}} & \multirow{2}{*}{\begin{tabular}[c]{@{}c@{}}Performance over \\ the best Partitioner\end{tabular}} \\ \cline{3-8}
\multicolumn{2}{|c|}{}                                                                              & \textsc{Cuttana}      & \textsc{Fennel} & \textsc{Ldg} & \multicolumn{1}{c|}{\textsc{HeiStream}} & \textsc{Hdrf} & \textsc{Ginger} &                                                                                                          &                                                                                                   \\ \hline \hline
\multicolumn{1}{|c|}{\multirow{3}{*}{twitter}}                        & PR                          & \cellcolor{blue!15}\textbf{168} & 813    & 811 & \multicolumn{1}{c|}{488}       & 413  & 492    & 66\%                                                                                                     & 59\%                                                                                              \\
\multicolumn{1}{|c|}{}                                                & CC                          & \cellcolor{blue!15}\textbf{33}  & 76     & 80  & \multicolumn{1}{c|}{81}        & 70   & 108    & 57\%                                                                                                     & 53\%                                                                                              \\
\multicolumn{1}{|c|}{}                                                & SSSP                        & \cellcolor{blue!15}\textbf{42}  & 176    & 202 & \multicolumn{1}{c|}{117}       & 81   & 77     & 64\%                                                                                                     & 45\%                                                                                              \\ \hline
\multicolumn{1}{|c|}{\multirow{3}{*}{uk07}}                           & PR                          & \cellcolor{blue!15}\textbf{141} & 293    & 336 & \multicolumn{1}{c|}{193}       & 1227 & 1269   & 27\%                                                                                                     & 27\%                                                                                              \\
\multicolumn{1}{|c|}{}                                                & CC                          & \cellcolor{blue!15}\textbf{63}  & 86     & 96  & \multicolumn{1}{c|}{84}        & 419  & 548    & 25\%                                                                                                     & 25\%                                                                                              \\
\multicolumn{1}{|c|}{}                                                & SSSP                        & \cellcolor{blue!15}\textbf{49}  & 61     & 63  & \multicolumn{1}{c|}{54}        & 181  & 147    & 9\%                                                                                                      & 9\%                                                                                               \\ \hline
\multicolumn{1}{|c|}{\multirow{3}{*}{rMat-XL}}                           & PR                          & \cellcolor{blue!15}\textbf{144} & 514    & 576 & \multicolumn{1}{c|}{376}       & 205  & 430    & 61\%                                                                                                     & 29\%                                                                                              \\
\multicolumn{1}{|c|}{}                                                & CC                          & \cellcolor{blue!15}\textbf{61}  & 93     & 112 & \multicolumn{1}{c|}{86}        & 74   & 82     & 29\%                                                                                                     & 25\%                                                                                              \\
\multicolumn{1}{|c|}{}                                                & SSSP                        & \cellcolor{blue!15}\textbf{53}  & 89     & 97  & \multicolumn{1}{c|}{65}        & 59   & 81     & 18\%                                                                                                     & 10\%                                                                                              \\ \hline
\end{tabular}
\end{table*}
    
\end{landscape}

We conduct case studies focusing on distributed graph analytics and graph databases to investigate how enhancements in quality metrics impact the performance metrics of these applications (e.g., throughput and execution time). We develop our application study framework on top of the benchmarking framework provided by Pacaci and Ozsu \cite{pacaci2019experimental}. We report performance metrics such as throughput and latency, but due to space constraints, we refer the reader to the original work for more information about the specifications \cite{pacaci2019experimental}.

\textbf{Distributed Graph Analytics.} Table \ref{table:analytic} shows the results of running three different algorithms on a Powerlyra cluster with 16 machines \cite{chen2019powerlyra} to assess the performance of various partitioning schemes on graph analytics to answer \textbf{RQ4}. We ran PageRank for 30 iterations, and connected components until we found all connected components, and single-source shortest path from a random vertex. We report the average latency of three runs. In practice, an algorithm can be executed multiple times (e.g., finding the shortest path from multiple sources in graphs with millions of vertices), which further increases the total latency improvement by using \textsc{Cuttana}. We show results for only the three largest graphs since small graphs can be processed more efficiently on a single machine than on multiple machines, because the network overhead introduced by adding a machine is not amortized by the parallelization achieved \cite{mcsherry2015scalability}.

In social network graphs, vertex partitioners other than \textsc{Cuttana}, suffer from significant load imbalance, overshadowing any advantages in network usage. This aligns with findings reported by Pacaci \cite{pacaci2019experimental}. We illustrate the reason for this imbalance in Figure \ref{fig:ebshow}. We used the edge-balance version of \model and the original implementation of other baselines. In web graphs, the worker imbalance was less pronounced, and vertex partitioners with better network overhead outperformed edge partitioners due to different message-passing protocols and low replication of low-degree vertices \cite{pacaci2019experimental}. More performance metrics are in Figure \ref{fig:crown}.

Among all algorithms, we observed the greatest performance improvement in PageRank. In this algorithm, most vertices are active in all iterations, stressing the system's network. \textsc{Cuttana}, which both balances edges like edge partitioners do and exploits the lower network overhead of vertex partitioners, provides a ``best-of-both-worlds'' choice. \model achieves our goal of producing sufficiently high-quality partitioning for large graphs that it improves runtime performance in analytical workloads. 


\textbf{Distributed Graph Database.} We conducted the LDBC social network benchmark \cite{erling2015ldbc} on a JanusGraph\footnote{We compare with vertex partitioners as JanusGraph requires the data to be vertex partitioned \cite{pacaci2019experimental}. } cluster of 4 machines with 24 concurrent client threads, using Cassandra as the backend storage engine. Our goal is to observe how improvements in edge-cut and edge imbalance can translate into improved throughput in a distributed graph database, addressing \textbf{RQ5}. The queries and graph were generated by the LDBC generator. The improvement in graph databases is smaller than in analytics, because existing partitioners produce less edge imbalance on the LDBC-generated graph than we observed on other graphs, e.g., Twitter. Additionally, LDBC one-hop and two-hop queries limit the number of returned neighbors, which we believe puts the system under less stress, and many queries can be answered locally, although there exist neighbors in other partitions (machines). We observed a 23\% improvement in the throughput of two-hop queries and a 7\% improvement for one-hop queries, without a major difference in tail latency. 

\begin{table}[]
\centering
\caption{Partitioning Metrics and Performance (throughput in the unit of query per second) of one-hop and two-hop neighborhood retrieval on LDBC social network benchmark.}
\label{table:graphdb}
\begin{tabular}{|c|cccc|}
\hline 
Metric           & \model & \fennel & \heistream & \ldg   \\ \hline \hline
Edge-cut         & 37.49   & 47.72  & 53.26     & 74.22 \\
Edge-imbalance   & 1.09    & 1.13   & 1.8       & 1.89  \\
Vertex-imbalance & 1.03    & 1.00   & 1.05      & 1.05  \\ \hline \hline
one-hop  (q/s)     & \cellcolor{blue!15}\textbf{2776}    & 2595   & 2381      & 1998  \\
two-hop (q/s)      & \cellcolor{blue!15}\textbf{232}     & 189    & 164       & 131   \\ \hline
\end{tabular}
\end{table}

\section{Related Work}

Distributed graph analytics has gained significant attention 
\cite{gonzalez2012powergraph,chen2019powerlyra,GraphX,fan2021graphscope,li2019topox,yan2020g,iyer2021tegra,wang2022scaleg,chen2023khuzdul} since the introduction of Pregel \cite{malewicz2010pregel}.
Many partitioning strategies have been proposed to reduce network overhead and address load imbalance \cite{tsourakakis2014fennel,stanton2012streaming,petroni2015hdrf,clugp,spnl,hpcd,mayer2018adwise,awadelkarim2020prioritized,chen2019powerlyra}, since partitioning plays a crucial role in application latency \cite{pacaci2019experimental}. Most of the recent advances in graph partitioning are in edge partitioning, which is unsurprising, since edge partitioners produce better edge balance than do vertex partitioners, and edge balance leads to even load distribution. \model takes a different approach and imposes an edge-balance condition while partitioning by vertex.
\textsc{Hdrf} \cite{petroni2015hdrf} and \textsc{Ginger} \cite{chen2019powerlyra} are two popular partitioners that reduce vertex replication and exhibit the best performance on large graphs \cite{pacaci2019experimental}. \textsc{Clugp} \cite{clugp} and \textsc{Hpcd} \cite{hpcd} are more recent edge-partitioners. \textsc{Clugp} provides a fast restreaming partitioning solution, while \textsc{Hpcd} transforms the problem into a combinatorial design problem.

However, some systems require vertex-partitioning \cite{pacaci2019experimental,faraj2022buffered}, and the message-passing protocol of the system changes when the graph is vertex-partitioned \cite{pacaci2019experimental}. Stanton et al. analyzed multiple scoring functions for streaming vertex partitioners and proposed \ldg \cite{stanton2012streaming}. Later, \fennel \cite{tsourakakis2014fennel} introduced a new scoring function with the same greedy, score-based model that outperformed \ldg and remained state-of-the-art for an extended period. \heistream \cite{faraj2022buffered} and \textsc{Spnl} \cite{spnl} are recent streaming partitioners whose evaluations showed it to be better than \fennel. We found that \model outperforms both \fennel and \heistream, especially on large graphs, which is the most compelling use case for streaming partitioners. Unfortunately, the code for \textsc{Spnl} was not available. However, in the common datasets UK07 and UK02, in both edge-balance and vertex-balance modes, our model exhibited better partitioning quality than that reported by \textsc{Spnl}.


Finally, there are in-memory partitioners for both edge- and vertex-partitioning \cite{sanders2011engineering,pellegrini1996scotch,karypis1998fast,zhang2017graph}. \textsc{Metis} \cite{karypis1998fast} is considered the gold standard for vertex-partitioning and \textsc{Ne} \cite{zhang2017graph} for edge-partitioning. In-memory partitioners inspired our coarsening and refinement strategy, but we adopted a different approach to facilitate scalability. 
In-memory partitioners offer better quality than streaming partitioners in medium-sized graphs; however, they fail to partition billion-scale graphs \cite{margo2017sorting,mayer2021hybrid,wang2023lightweight}.
\textsc{Kl} \cite{kernighan1970efficient} and \textsc{Fm} \cite{fiduccia1988linear} are partitioning methods based on vertex exchange.
\model differs from both of these approaches.
First, it includes a coarsening phase to efficiently reduce graph size. Our coarsening approach also differs from that of \textsc{Metis}, which relies on multiple maximal matching iterations, rendering it unscalable for large graphs. We reframe coarsening as another streaming partitioning problem (sub-partitioning).
Second, at the core of refinement, \textsc{Kl} swaps vertices while we move subpartitions; our approach provides asymptotically better performance.
While the greedy heuristics and moves in \textsc{Fm} are similar to \textsc{Cuttana}'s, its bucket listing technique does not apply to our case as it only works for small unweighted graphs. \textsc{Fm} requires $\K^2$ buckets for $\K$-way partitioning, each sized by the quality gain, making it unscalable for our weighted graph with gains up to $O(\frac{V^2}{\K})$. Finally, we initiate refinement on a graph partitioned by a streaming partitioner, which converges more quickly. This facilitates parallel coarsening and data structure preprocessing during streaming, thereby enhancing overall efficiency.
Another approach to improving the partitioning quality of streaming partitioners is restreaming \cite{awadelkarim2020prioritized,nishimura2013restreaming}, where the graph is read multiple times to iteratively improve partitioning quality. The restreaming technique is orthogonal to our work, and \model can be used in a restreaming system for faster convergence.  There are also distributed graph partitioners that improve runtime and memory constraints over single-node solutions \cite{hanai2019distributed,slota2020scalable,martella2017spinner,margo2015scalable}. 

In graph databases, updates can degrade partition quality over time. \textsc{Cuttana} can be combined with incremental graph partitioning techniques, such as those of \textsc{Leopard} \cite{huang2016leopard} and Fan et al \cite{fan2020incrementalization} to work in the dynamic graph setting. Another possibility is a periodic coarse-grained repartitioning of the entire graph or fine-grained recalculation of the scoring function to determine when to move vertices. Repartitioning can be performed in the background, and its overhead can be negligible in long-running applications.

Since the choice of the best partitioner varies based on the dataset, Merkel proposed a machine learning model to select the best partitioner based on workload features \cite{merkel2023partitioner}. However, \model shows robust performance improvement in large datasets in both web and social domains. The trend in the development of distributed applications for graphs has not stopped and has recently been accelerated by the development of distributed systems for graph learning \cite{md2021distgnn,wang2021gnnadvisor,vatter2023evolution,zheng2020distdgl,gandhi2021p3}.

\section{Limitations, Conclusion, and Future Work}
\textbf{Limitations.}
Using \textsc{Cuttana} for dynamic graphs requires repartitioning or incorporating existing incremental approaches \cite{fan2020incrementalization, huang2016leopard}, which we have not yet undertaken. Using \textsc{Cuttana} for small and mid-scale graphs, such as Orkut, may not be a good choice in single-run analytics, as the performance gain in analytical job runtime will not amortize the partitioning latency. Because our additional overhead compared to streaming solutions is independent of graph size, \textsc{Cuttana}'s sweet spot is large graphs where the additional overhead is small relative to streaming time.
\textsc{Cuttana} is designed to be a partitioner for distributed vertex-centric applications on massive graphs; in-memory solutions can provide better partitioning for small-to-medium scale graphs with higher quality.

\textbf{Future Work \& Conclusion.} We introduced a novel streaming partitioner that incorporates prioritized buffering to improve the quality of classic streaming graph partitioners. We then conceptualized the problem of improving the initial partitioning by relocating vertices and presented a coarsening and refinement strategy capable of improving the quality of the initial output of any partitioner. The refinement algorithm demonstrated theoretical efficiency, with time complexity independent of graph size. Our partitioner, \textsc{Cuttana}, significantly improved the partitioning quality of its core streaming counterpart, surpassing state-of-the-art vertex partitioners in various scenarios, considering different quality metrics and balance conditions.
With a parallel implementation and leveraging the power-law property of large graphs,
\model's parallel implementation incurs negligible partitioning latency overhead relative to a simple streaming partitioner. Our application study confirmed that using \model almost always leads to lower network overhead and even load distribution, resulting in better runtimes and throughputs for both graph analytics and database applications. Consequently, \model emerges as the preferred option for graph partitioning. Looking ahead, we envision further advances in the form of developing a new scoring function for buffering and extending our generalized trade concept to address more complex moves. In cases where moving a single sub-partition fails to enhance quality, however, relocating two or more vertices simultaneously can maintain the balance condition and improve overall quality. 

\newpage

\bibliographystyle{acm} 
\bibliography{sample} 

\begin{thebibliography}{10}

\bibitem{GraphX}
Graphx: https://spark.apache.org/graphx/.

\bibitem{JanusGraph}
Janusgraph: https://janusgraph.org/.

\bibitem{lfqueue}
Lock-free queue: https://github.com/cameron314/readerwriterqueue.

\bibitem{TitanDB}
Titan db: https://titan.thinkaurelius.com/.

\bibitem{usroad}
Us-road-dataset: https://networkrepository.com/road-road-usa.php.

\bibitem{abbas2018streaming}
{\sc Abbas, Z., Kalavri, V., Carbone, P., and Vlassov, V.}
\newblock Streaming graph partitioning: an experimental study.
\newblock {\em Proceedings of the VLDB Endowment 11}, 11 (2018), 1590--1603.

\bibitem{albert2000error}
{\sc Albert, R., Jeong, H., and Barab{\'a}si, A.-L.}
\newblock Error and attack tolerance of complex networks.
\newblock {\em nature 406}, 6794 (2000), 378--382.

\bibitem{awadelkarim2020prioritized}
{\sc Awadelkarim, A., and Ugander, J.}
\newblock Prioritized restreaming algorithms for balanced graph partitioning.
\newblock In {\em Proceedings of the 26th ACM SIGKDD International Conference on Knowledge Discovery \& Data Mining\/} (2020), pp.~1877--1887.

\bibitem{bourse2014balanced}
{\sc Bourse, F., Lelarge, M., and Vojnovic, M.}
\newblock Balanced graph edge partition.
\newblock In {\em Proceedings of the 20th ACM SIGKDD international conference on Knowledge discovery and data mining\/} (2014), pp.~1456--1465.

\bibitem{buragohain2020a1}
{\sc Buragohain, C., Risvik, K.~M., Brett, P., Castro, M., Cho, W., Cowhig, J., Gloy, N., Kalyanaraman, K., Khanna, R., Pao, J., et~al.}
\newblock A1: A distributed in-memory graph database.
\newblock In {\em Proceedings of the 2020 ACM SIGMOD International Conference on Management of Data\/} (2020), pp.~329--344.

\bibitem{ccatalyurek2023more}
{\sc {\c{C}}ataly{\"u}rek, {\"U}., Devine, K., Faraj, M., Gottesb{\"u}ren, L., Heuer, T., Meyerhenke, H., Sanders, P., Schlag, S., Schulz, C., Seemaier, D., et~al.}
\newblock More recent advances in (hyper) graph partitioning.
\newblock {\em ACM Computing Surveys 55}, 12 (2023), 1--38.

\bibitem{chen2023khuzdul}
{\sc Chen, J., and Qian, X.}
\newblock Khuzdul: Efficient and scalable distributed graph pattern mining engine.
\newblock In {\em Proceedings of the 28th ACM International Conference on Architectural Support for Programming Languages and Operating Systems, Volume 2\/} (2023), pp.~413--426.

\bibitem{chen2019powerlyra}
{\sc Chen, R., Shi, J., Chen, Y., Zang, B., Guan, H., and Chen, H.}
\newblock Powerlyra: Differentiated graph computation and partitioning on skewed graphs.
\newblock {\em ACM Transactions on Parallel Computing (TOPC) 5}, 3 (2019), 1--39.

\bibitem{de2000computational}
{\sc De~Berg, M.}
\newblock {\em Computational geometry: algorithms and applications}.
\newblock Springer Science \& Business Media, 2000.

\bibitem{erling2015ldbc}
{\sc Erling, O., Averbuch, A., Larriba-Pey, J., Chafi, H., Gubichev, A., Prat, A., Pham, M.-D., and Boncz, P.}
\newblock The ldbc social network benchmark: Interactive workload.
\newblock In {\em Proceedings of the 2015 ACM SIGMOD International Conference on Management of Data\/} (2015), pp.~619--630.

\bibitem{fan2021graphscope}
{\sc Fan, W., He, T., Lai, L., Li, X., Li, Y., Li, Z., Qian, Z., Tian, C., Wang, L., Xu, J., et~al.}
\newblock Graphscope: a unified engine for big graph processing.
\newblock {\em Proceedings of the VLDB Endowment 14}, 12 (2021), 2879--2892.

\bibitem{fan2020incrementalization}
{\sc Fan, W., Liu, M., Tian, C., Xu, R., and Zhou, J.}
\newblock Incrementalization of graph partitioning algorithms.
\newblock {\em Proceedings of the VLDB Endowment 13}, 8 (2020), 1261--1274.

\bibitem{fan2023application}
{\sc Fan, W., Xu, R., Yin, Q., Yu, W., and Zhou, J.}
\newblock Application-driven graph partitioning.
\newblock {\em The VLDB Journal 32}, 1 (2023), 149--172.

\bibitem{faraj2022buffered}
{\sc Faraj, M.~F., and Schulz, C.}
\newblock Buffered streaming graph partitioning.
\newblock {\em ACM Journal of Experimental Algorithmics 27\/} (2022), 1--26.

\bibitem{fiduccia1988linear}
{\sc Fiduccia, C.~M., and Mattheyses, R.~M.}
\newblock A linear-time heuristic for improving network partitions.
\newblock In {\em Papers on Twenty-five years of electronic design automation}. 1988, pp.~241--247.

\bibitem{gandhi2021p3}
{\sc Gandhi, S., and Iyer, A.~P.}
\newblock P3: Distributed deep graph learning at scale.
\newblock In {\em 15th $\{$USENIX$\}$ Symposium on Operating Systems Design and Implementation ($\{$OSDI$\}$ 21)\/} (2021), pp.~551--568.

\bibitem{garey1974some}
{\sc Garey, M.~R., Johnson, D.~S., and Stockmeyer, L.}
\newblock Some simplified np-complete problems.
\newblock In {\em Proceedings of the sixth annual ACM symposium on Theory of computing\/} (1974), pp.~47--63.

\bibitem{gonzalez2012powergraph}
{\sc Gonzalez, J.~E., Low, Y., Gu, H., Bickson, D., and Guestrin, C.}
\newblock $\{$PowerGraph$\}$: Distributed $\{$Graph-Parallel$\}$ computation on natural graphs.
\newblock In {\em 10th USENIX symposium on operating systems design and implementation (OSDI 12)\/} (2012), pp.~17--30.

\bibitem{hanai2019distributed}
{\sc Hanai, M., Suzumura, T., Tan, W.~J., Liu, E., Theodoropoulos, G., and Cai, W.}
\newblock Distributed edge partitioning for trillion-edge graphs.
\newblock {\em arXiv preprint arXiv:1908.05855\/} (2019).

\bibitem{huang2016leopard}
{\sc Huang, J., and Abadi, D.~J.}
\newblock Leopard: Lightweight edge-oriented partitioning and replication for dynamic graphs.
\newblock {\em Proceedings of the VLDB Endowment 9}, 7 (2016).

\bibitem{iyer2021tegra}
{\sc Iyer, A.~P., Pu, Q., Patel, K., Gonzalez, J.~E., and Stoica, I.}
\newblock $\{$TEGRA$\}$: Efficient $\{$Ad-Hoc$\}$ analytics on evolving graphs.
\newblock In {\em 18th USENIX Symposium on Networked Systems Design and Implementation (NSDI 21)\/} (2021), pp.~337--355.

\bibitem{karypis1998fast}
{\sc Karypis, G., and Kumar, V.}
\newblock A fast and high quality multilevel scheme for partitioning irregular graphs.
\newblock {\em SIAM Journal on scientific Computing 20}, 1 (1998), 359--392.

\bibitem{kernighan1970efficient}
{\sc Kernighan, B.~W., and Lin, S.}
\newblock An efficient heuristic procedure for partitioning graphs.
\newblock {\em The Bell system technical journal 49}, 2 (1970), 291--307.

\bibitem{wsvr}
{\sc Khorasani, F., Gupta, R., and Bhuyan, L.~N.}
\newblock Scalable simd-efficient graph processing on gpus.
\newblock In {\em Proceedings of the 24th International Conference on Parallel Architectures and Compilation Techniques\/} (2015), PACT '15, pp.~39--50.

\bibitem{kleinberg1999web}
{\sc Kleinberg, J.~M., Kumar, R., Raghavan, P., Rajagopalan, S., and Tomkins, A.~S.}
\newblock The web as a graph: Measurements, models, and methods.
\newblock In {\em Computing and Combinatorics: 5th Annual International Conference, COCOON’99 Tokyo, Japan, July 26--28, 1999 Proceedings 5\/} (1999), Springer, pp.~1--17.

\bibitem{clugp}
{\sc Kong, D., Xie, X., and Zhang, Z.}
\newblock Clustering-based partitioning for large web graphs.
\newblock In {\em 2022 IEEE 38th International Conference on Data Engineering (ICDE)\/} (2022), IEEE, pp.~593--606.

\bibitem{konect}
{\sc Kunegis, J.}
\newblock {KONECT} -- {The} {Koblenz} {Network} {Collection}.
\newblock In {\em Proc. Int. Conf. on World Wide Web Companion\/} (2013), pp.~1343--1350.

\bibitem{li2022bytegraph}
{\sc Li, C., Chen, H., Zhang, S., Hu, Y., Chen, C., Zhang, Z., Li, M., Li, X., Han, D., Chen, X., et~al.}
\newblock Bytegraph: a high-performance distributed graph database in bytedance.
\newblock {\em Proceedings of the VLDB Endowment 15}, 12 (2022), 3306--3318.

\bibitem{li2019topox}
{\sc Li, D., Zhang, Y., Wang, J., and Tan, K.-L.}
\newblock Topox: Topology refactorization for efficient graph partitioning and processing.
\newblock {\em Proceedings of the VLDB Endowment 12}, 8 (2019), 891--905.

\bibitem{malewicz2010pregel}
{\sc Malewicz, G., Austern, M.~H., Bik, A.~J., Dehnert, J.~C., Horn, I., Leiser, N., and Czajkowski, G.}
\newblock Pregel: a system for large-scale graph processing.
\newblock In {\em Proceedings of the 2010 ACM SIGMOD International Conference on Management of data\/} (2010), pp.~135--146.

\bibitem{margo2015scalable}
{\sc Margo, D., and Seltzer, M.}
\newblock A scalable distributed graph partitioner.
\newblock {\em Proceedings of the VLDB Endowment 8}, 12 (2015), 1478--1489.

\bibitem{margo2017sorting}
{\sc Margo, D.~W.}
\newblock {\em Sorting Shapes the Performance of Graph-Structured Systems}.
\newblock PhD thesis, Harvard University, 2017.

\bibitem{martella2017spinner}
{\sc Martella, C., Logothetis, D., Loukas, A., and Siganos, G.}
\newblock Spinner: Scalable graph partitioning in the cloud.
\newblock In {\em 2017 IEEE 33rd international conference on data engineering (ICDE)\/} (2017), Ieee, pp.~1083--1094.

\bibitem{mayer2018adwise}
{\sc Mayer, C., Mayer, R., Tariq, M.~A., Geppert, H., Laich, L., Rieger, L., and Rothermel, K.}
\newblock Adwise: Adaptive window-based streaming edge partitioning for high-speed graph processing.
\newblock In {\em 2018 IEEE 38th International Conference on Distributed Computing Systems (ICDCS)\/} (2018), IEEE, pp.~685--695.

\bibitem{mayer2021hybrid}
{\sc Mayer, R., and Jacobsen, H.-A.}
\newblock Hybrid edge partitioner: Partitioning large power-law graphs under memory constraints.
\newblock In {\em Proceedings of the 2021 International Conference on Management of Data\/} (2021), pp.~1289--1302.

\bibitem{mccune2015thinking}
{\sc McCune, R.~R., Weninger, T., and Madey, G.}
\newblock Thinking like a vertex: a survey of vertex-centric frameworks for large-scale distributed graph processing.
\newblock {\em ACM Computing Surveys (CSUR) 48}, 2 (2015), 1--39.

\bibitem{mcsherry2015scalability}
{\sc McSherry, F., Isard, M., and Murray, D.~G.}
\newblock Scalability! but at what $\{$COST$\}$?
\newblock In {\em 15th Workshop on Hot Topics in Operating Systems (HotOS XV)\/} (2015).

\bibitem{md2021distgnn}
{\sc Md, V., Misra, S., Ma, G., Mohanty, R., Georganas, E., Heinecke, A., Kalamkar, D., Ahmed, N.~K., and Avancha, S.}
\newblock Distgnn: Scalable distributed training for large-scale graph neural networks.
\newblock In {\em Proceedings of the International Conference for High Performance Computing, Networking, Storage and Analysis\/} (2021), pp.~1--14.

\bibitem{merkel2023partitioner}
{\sc Merkel, N., Mayer, R., Fakir, T.~A., and Jacobsen, H.-A.}
\newblock Partitioner selection with ease to optimize distributed graph processing.
\newblock In {\em 2023 IEEE 39th International Conference on Data Engineering (ICDE)\/} (2023), IEEE.

\bibitem{nishimura2013restreaming}
{\sc Nishimura, J., and Ugander, J.}
\newblock Restreaming graph partitioning: simple versatile algorithms for advanced balancing.
\newblock In {\em Proceedings of the 19th ACM SIGKDD international conference on Knowledge discovery and data mining\/} (2013), pp.~1106--1114.

\bibitem{pacaci2019experimental}
{\sc Pacaci, A., and {\"O}zsu, M.~T.}
\newblock Experimental analysis of streaming algorithms for graph partitioning.
\newblock In {\em Proceedings of the 2019 International Conference on Management of Data\/} (2019), pp.~1375--1392.

\bibitem{pellegrini1996scotch}
{\sc Pellegrini, F., and Roman, J.}
\newblock Scotch: A software package for static mapping by dual recursive bipartitioning of process and architecture graphs.
\newblock In {\em High-Performance Computing and Networking: International Conference and Exhibition HPCN EUROPE 1996 Brussels, Belgium, April 15--19, 1996 Proceedings 4\/} (1996), Springer, pp.~493--498.

\bibitem{petroni2015hdrf}
{\sc Petroni, F., Querzoni, L., Daudjee, K., Kamali, S., and Iacoboni, G.}
\newblock Hdrf: Stream-based partitioning for power-law graphs.
\newblock In {\em Proceedings of the 24th ACM international on conference on information and knowledge management\/} (2015), pp.~243--252.

\bibitem{hpcd}
{\sc Qu, W., Zhang, W., Cheng, J., Zhang, C., Han, W., Bai, B., Zhang, C.~J., He, L., and Wang, X.}
\newblock Optimizing graph partition by optimal vertex-cut: A holistic approach.
\newblock In {\em 2023 IEEE 39th International Conference on Data Engineering (ICDE)\/} (2023), IEEE, pp.~1019--1031.

\bibitem{sahu2017ubiquity}
{\sc Sahu, S., Mhedhbi, A., Salihoglu, S., Lin, J., and {\"O}zsu, M.~T.}
\newblock The ubiquity of large graphs and surprising challenges of graph processing.
\newblock {\em Proceedings of the VLDB Endowment 11}, 4 (2017), 420--431.

\bibitem{sanders2011engineering}
{\sc Sanders, P., and Schulz, C.}
\newblock Engineering multilevel graph partitioning algorithms.
\newblock In {\em European Symposium on algorithms\/} (2011), Springer, pp.~469--480.

\bibitem{slota2020scalable}
{\sc Slota, G.~M., Root, C., Devine, K., Madduri, K., and Rajamanickam, S.}
\newblock Scalable, multi-constraint, complex-objective graph partitioning.
\newblock {\em IEEE Transactions on Parallel and Distributed Systems 31}, 12 (2020), 2789--2801.

\bibitem{stanton2012streaming}
{\sc Stanton, I., and Kliot, G.}
\newblock Streaming graph partitioning for large distributed graphs.
\newblock In {\em Proceedings of the 18th ACM SIGKDD international conference on Knowledge discovery and data mining\/} (2012), pp.~1222--1230.

\bibitem{tsourakakis2014fennel}
{\sc Tsourakakis, C., Gkantsidis, C., Radunovic, B., and Vojnovic, M.}
\newblock Fennel: Streaming graph partitioning for massive scale graphs.
\newblock In {\em Proceedings of the 7th ACM international conference on Web search and data mining\/} (2014), pp.~333--342.

\bibitem{valois1994implementing}
{\sc Valois, J.~D.}
\newblock Implementing lock-free queues.
\newblock In {\em Proceedings of the seventh international conference on Parallel and Distributed Computing Systems\/} (1994), Citeseer, pp.~64--69.

\bibitem{vatter2023evolution}
{\sc Vatter, J., Mayer, R., and Jacobsen, H.-A.}
\newblock The evolution of distributed systems for graph neural networks and their origin in graph processing and deep learning: A survey.
\newblock {\em ACM Computing Surveys\/} (2023).

\bibitem{wang2022scaleg}
{\sc Wang, X., Wen, D., Qin, L., Chang, L., and Zhang, W.}
\newblock Scaleg: A distributed disk-based system for vertex-centric graph processing.
\newblock In {\em 2022 IEEE 38th International Conference on Data Engineering (ICDE)\/} (2022), IEEE, pp.~1511--1512.

\bibitem{wang2021gnnadvisor}
{\sc Wang, Y., Feng, B., Li, G., Li, S., Deng, L., Xie, Y., and Ding, Y.}
\newblock $\{$GNNAdvisor$\}$: An adaptive and efficient runtime system for $\{$GNN$\}$ acceleration on $\{$GPUs$\}$.
\newblock In {\em 15th USENIX symposium on operating systems design and implementation (OSDI 21)\/} (2021), pp.~515--531.

\bibitem{spnl}
{\sc Wang, Z., Yang, Z., Wang, N., Du, Y., Nie, J., Wei, Z., Gu, Y., and Yu, G.}
\newblock Lightweight streaming graph partitioning by fully utilizing knowledge from local view.
\newblock In {\em 2023 IEEE 43rd International Conference on Distributed Computing Systems (ICDCS)\/} (2023), IEEE, pp.~614--625.

\bibitem{wang2023lightweight}
{\sc Wang, Z., Yang, Z., Wang, N., Du, Y., Nie, J., Wei, Z., Gu, Y., and Yu, G.}
\newblock Lightweight streaming graph partitioning by fully utilizing knowledge from local view.
\newblock In {\em 2023 IEEE 43rd International Conference on Distributed Computing Systems (ICDCS)\/} (2023), IEEE, pp.~614--625.

\bibitem{yan2020g}
{\sc Yan, D., Guo, G., Chowdhury, M. M.~R., {\"O}zsu, M.~T., Ku, W.-S., and Lui, J.~C.}
\newblock G-thinker: A distributed framework for mining subgraphs in a big graph.
\newblock In {\em 2020 IEEE 36th International Conference on Data Engineering (ICDE)\/} (2020), IEEE, pp.~1369--1380.

\bibitem{zhang2017graph}
{\sc Zhang, C., Wei, F., Liu, Q., Tang, Z.~G., and Li, Z.}
\newblock Graph edge partitioning via neighborhood heuristic.
\newblock In {\em Proceedings of the 23rd ACM SIGKDD International Conference on Knowledge Discovery and Data Mining\/} (2017), pp.~605--614.

\bibitem{zheng2020distdgl}
{\sc Zheng, D., Ma, C., Wang, M., Zhou, J., Su, Q., Song, X., Gan, Q., Zhang, Z., and Karypis, G.}
\newblock Distdgl: distributed graph neural network training for billion-scale graphs.
\newblock In {\em 2020 IEEE/ACM 10th Workshop on Irregular Applications: Architectures and Algorithms (IA3)\/} (2020), IEEE, pp.~36--44.

\end{thebibliography}

\end{document}